\documentclass{aa}  
\usepackage{natbib}

\usepackage{graphicx}

\usepackage{txfonts}

\usepackage{hyperref}

\begin{document}

     \title{The (limited) effect of viscosity in multiphase turbulent mixing}

   \author{Tirso Marin-Gilabert\inst{\ref{USM}, \ref{MPA},\ref{CfA}}\thanks{tmarin@mpa-garching.mpg.de} 
   \and Max Gronke \inst{\ref{ARI}, \ref{MPA}}
   \and S. Peng Oh \inst{\ref{UCSB}}}
   
   \institute{Universitäts-Sternwarte, Fakultät für Physik, Ludwig-Maximilians-Universität München, Scheinerstr. 1, 81679 München, Germany\label{USM} \and 
    Centre for Astronomy of Heidelberg University, Astronomisches Rechen-Institut, Mönchhofstr. 12-14, 69120 Heidelberg, Germany\label{ARI} \and
   Max Planck Institute for Astrophysics, Garching 85748, Germany \label{MPA} \and
   Center for Astrophysics | Harvard \& Smithsonian, 60 Garden St. Cambridge, MA 02138, USA \label{CfA} \and
   Department of Physics, University of California, Santa Barbara, CA 93106, USA \label{UCSB}}

   \date{Received -; accepted -}

  \abstract 
   {Multiphase gas can be found in many astrophysical environments, such as galactic outflows, stellar wind bubbles, and the circumgalactic medium, where the interplay between turbulence, cooling, and viscosity can significantly influence gas dynamics and star formation processes. We investigate the role of viscosity in modulating turbulence and radiative cooling in turbulent radiative mixing layers (TRMLs). In particular, we aim to determine how different amounts of viscosity affect the Kelvin-Helmholtz instability (KHI), turbulence evolution, and the efficiency of gas mixing and cooling. Using idealized 2D numerical setups, we computed the critical viscosity required to suppress the KHI in shear flows characterized by different density contrasts and Mach numbers. These results were then used in a 3D shear layer setup to explore the impact of viscosity on cooling efficiency and turbulence across different cooling regimes. We find that the critical viscosity follows the expected dependence on overdensity and Mach number. Our viscous TRML simulations show different behaviors in the weak and strong cooling regimes. In the weak cooling regime, viscosity has a strong impact, resulting in laminar flows and breaking previously established inviscid relations between cooling and turbulence (albeit leaving the total luminosity unaffected). However, in the strong cooling regime, where cooling timescales are shorter than viscous timescales, key scaling relations in TRMLs remain largely intact. In this regime, which must hold for gas to remain multiphase, radiative losses dominate, and the system effectively behaves as nonviscous regardless of the actual level of viscosity. Our findings have direct implications for the interpretation of observational diagnostics and the development of subgrid models in large-scale simulations.}


   \maketitle

\section{Introduction} \label{sec:introduction}

Turbulence and multiphase systems are ubiquitous in plenty of astrophysical environments, including the interstellar medium \citep[ISM;][]{McKee_1977, Audit_2010}, circumgalactic medium \citep[CGM;][]{Tumlinson_2017, Faucher_2023}, galactic winds \citep{Veilleux_2020,Thompson_2024}, and intracluster medium \citep[ICM;][]{Markevitch_2007, Kravtsov_2012}. Therefore, a proper modeling of turbulence and multiphase gas is fundamental to understanding key processes in astrophysics.

Observations provide clear evidence of multiphase gas in the CGM, as well as in other astrophysical environments. For instance, hot gas ($T \gtrsim 10^6$~K) in halos around galaxies has been detected in X-rays \citep[e.g.,][]{Anderson_2010, Bregman_2018}. Meanwhile, cool gas ($T \sim 10^4$~K) has been detected through Ly$\alpha$ emission \citep[e.g.,][]{Steidel_2010, Wisotzki_2018, Arrigoni_2023}, as well as absorption lines in the spectra of background quasars \citep[e.g.,][]{Tripp_1998, Peroux_2019}.

Turbulence is inherent to astrophysics, influencing the gas dynamics across very diverse environments. It plays a crucial role in distributing the energy from large to small scales in a turbulent cascade. The energy is injected at large scales and it cascades down until it reaches the Kolmogorov scale, where the kinetic energy is dissipated into heat \citep{Kolmogorov_1941, Kolmogorov_1962}. Turbulence is key across many different scales in astrophysics: from small scales such as stellar interiors \citep{Bruggen_2001} and protoplanetary disks \citep{Johansen_2006} to large scales such as active galactic nucleus jets \citep{Scannapieco_2008} and cold fronts \citep{Zuhone_2010}.

Turbulence mixes the hot and cold gas, producing intermediate temperature gas. In nonideal gases, this intermediate gas is unstable and determines the amount of cold gas available for star formation. This mass transfer has been extensively studied in the context of turbulent radiative mixing layers \citep[TRMLs; e.g.,][]{Begelman_1990, Slavin_1993}, which can be applied to several astrophysical systems, such as galactic outflows \citep{Bruggen_2016, Schneider_2017, Fielding_2022} or stellar wind bubbles \citep{El-Badry_2019, Lancaster_2021a, Lancaster_2021b}. Based on combustion theory, \cite{Tan_2021} identified a key relation between the turbulent and cooling timescales, expressed by the Damköhler number: ${\rm Da} = t_{\rm turb} / t_{\rm cool}$ \citep[][]{Damkohler_1940}. This dimensionless number characterizes the cooling strength and describes a different behavior of cooling efficiency, depending on whether the system is in the weak or the strong cooling regime. Roughly speaking, when ${\rm Da} < 1$, cooling is slow compared to turbulence and mixing dominates the gas entropy. When ${\rm Da} > 1,$ cooling is faster than turbulence and  the mixed gas is cooled down rapidly, leading to a multiphase system.

The total bolometric luminosity of a system depends on how much gas can be mixed to  intermediate temperatures; therefore, the suppression of turbulence and mixing might lead to differences in luminosity. This has been studied in the context of TRMLs with magnetic fields, where the magnetic field orientation is key for the suppression of the Kelvin-Helmholtz instability (KHI; \citealt{Chandrasekhar_1961}). \cite{Ji_2019} found that regardless of the initial orientation, magnetic fields are amplified by the shear motions, leading to the suppression of mixing and, therefore, less cooling, in the slow cooling regime. However, \cite{Das_2023} found that although magnetic fields suppress mixing,  the relation between turbulent velocity and cooling rate found by \cite{Tan_2021} in the
fast cooling regime still holds in the magnetized case.

Transport processes such as viscosity also affect the development of instabilities and mixing. Even though they are non-magnetized, fully ionized plasmas are expected to be very viscous \citep{Spitzer_1962, Zeldovich_1967} and the effective viscosity is reduced in the presence of magnetic fields \citep{Braginskii_1965, Squire_2023} and plasma microinstabilities \citep{Schekochihin_2006, Kunz_2014}. Therefore, the overall effect of viscosity in astrophysical systems is still under debate \citep{Zhuravleva_2019, Heinrich_2024, Marin-Gilabert_2024}. Although the net effective viscosity in astrophysical systems is not clear yet, it definitely influences the development of instabilities and turbulence \citep{Roediger_2013, Marin-Gilabert_2022}. In a similar way to magnetic fields, viscosity changes the mixing efficiency in multiphase systems. In this paper, we aim to study in detail these differences in turbulence. In particular, we consider whether this might impact the universal relation between turbulent velocity and cooling rate for inviscid TRMLs \citep{Fielding_2020, Tan_2021}.

Viscosity sets the Kolmogorov scale \citep[$\eta_{\rm L}$;][]{Kolmogorov_1941}, namely, the scale at which the kinetic energy is transformed into heat in the turbulent cascade. Below this scale, hydrodynamical instabilities are suppressed. On the other hand, conduction sets the Field length \citep[$\delta_{\rm L}$;][]{Field_1965}, the maximum scale over which conductive energy transport is effective in the presence of radiative cooling \citep{Begelman_1990b}. Thermal instabilities are suppressed on scales smaller than the Field length. An important dimensionless parameter in combustion theory is the Karlovitz number \citep{Klimov_1963, Williams_1975}, which compares these two scales, ${\rm Ka} = \delta_{\rm L}^2 / \eta_{\rm L}^2$. The Karlovitz number relates the flame thickness to the turbulent eddies scale, determining the shape of the flame front. If $\rm Ka < 1$, the small-scale turbulence is not small enough to break up the flame and the combustion happens in a laminar flow. If $\rm Ka > 1$, the small-scale turbulence grows faster than it can locally burn through its thickness, leading a turbulent flame front \citep{Kuo_2012}. Translated to TRMLs terminology, $\rm Ka < 1$ describes a scenario in which cooling takes place in a laminar regime, whereas if $\rm Ka > 1$, cooling happens in a turbulent regime. In the absence of explicit conduction and viscosity, the Kolmogorov scale and the scale of numerical heat diffusion are given by a similar multiple of the resolution scale in both cases ($\eta_{\rm L} \sim \delta_{\rm L} \sim \Delta x$) and, thus, $\rm Ka \sim 1$ \citep{Tan_2021}. In this paper, we study the parameter space where $\rm Ka < 1$ by explicitly including viscosity, while relying on numerical conduction, thus ensuring $\eta_{\rm L} > \delta_{\rm L}$.

The aim of this study was to understand how viscosity affects fluid dynamics in idealized setups, for both incompressible and compressible regimes. The findings of these control cases are then applied to TRMLs, where we study the interplay between viscosity and cooling. By including an explicit viscosity, we studied the parameter space in which $\rm Ka < 1$ and the implications that this might have for turbulence and mixing, as well as for cooling efficiency. In particular, we investigate whether the universal relations found for inviscid fluids \citep{Fielding_2020, Tan_2021} still hold. These results have potential implications in different astrophysical systems. For instance, they might influence the star formation rate in jellyfish galaxies or the gas dynamics in galactic outflows and the CGM.

This paper is organized as follows. In Sect. \ref{sec:theory} we introduce the relevant theory. The different numerical setups we use are explained in Sect. \ref{sec:numerical_setup}. In Sects. \ref{sec:crit_visc_overdens} and \ref{sec:crit_visc_mach} we show the results from idealized 2D simulations, while in Sect. \ref{sec:TRMLs} we show the results for TRMLs. We expand our discussion in Sect. \ref{sec:discussion} and present our conclusions in Sect. \ref{sec:conclusions}.

\section{Theoretical considerations} \label{sec:theory}

The fluid dynamics are described by the equations of hydrodynamics: conservation of mass, momentum and energy. Including viscosity, the equations are given by
\begin{equation}
    \frac{\partial \rho}{\partial t} + \nabla \cdot \left( \rho \vec{v} \right) = 0 \, ,
    \label{eqn:continuity_eq}
\end{equation}
\begin{equation}
    \frac{\partial (\rho \vec{v})}{\partial t} + \nabla \cdot \left(\rho \vec{v} \vec{v} + P\right) = -\nabla \cdot \vec{\sigma} \, ,
    \label{eqn:momentum_eq}
\end{equation}
\begin{equation}
     \frac{\partial (\rho e)}{\partial t} + \nabla \cdot \left[ \left(\rho e + P \right) \, \vec{v} \right] = -\nabla \cdot \left( \vec{v} \cdot \vec{\sigma} \right) \, .
     \label{eqn:energy_eq}
\end{equation}
$\rho$ is the gas density, $\vec{v}$ is the velocity of the gas, $P$ is the thermal pressure given by
\begin{equation}
    P = (\gamma - 1) \rho e_{\rm int} 
\end{equation}
and $e$ is the total energy per unit mass
\begin{equation}
    e = e_{\rm int} + \frac{1}{2} \vec{v}^2 \, ,
\end{equation}
where $e_{\rm int}$ is the internal energy per unit mass. Finally, $\vec{\sigma}$ is the isotropic viscous stress tensor
\begin{equation}
   \vec{\sigma} = \eta \, \left( \nabla \vec{v} + \nabla \vec{v}^{\rm T} - \frac{2}{3} \nabla \cdot \vec{v} \right), \, 
    \label{eqn:viscous_stress_tensor}
\end{equation}
where we have omitted the bulk viscosity term, which is zero for monoatomic gases, since it is related to the degree of freedom of molecular rotations \cite[see, e.g.,][]{Zeldovich_1967, Pitaevskii_1981}. Then, $\eta$ is the shear viscosity coefficient, which for fully ionized (non-magnetized) plasmas depends on the temperature of the gas as $\eta \propto T^{5/2}$ \citep{Spitzer_1962, Braginskii_1965}. For partially ionized plasmas, the effect of viscosity is reduced and depends on the ionized fraction \citep{Su_2017, Hopkins_2019}.

\subsection{Critical viscosity theoretical estimate} \label{sec:theoretical_estimate}

For reference, our numerical simulations will be computed as a function of the critical viscosity needed to suppress the growth of the KHI. Due to the nonlinear behavior of the KHI, there is no analytical solution to describe its evolution, so we need to estimate it based on a different set of assumptions. The theoretical estimate we used in this work comes from \citet{Roediger_2013} and later used in \citet{Marin-Gilabert_2022}. 

Viscosity smooths out the shear velocity gradient over a length scale $\pm d$ above and below the interface due to momentum diffusion. This leads to the suppression of the KHI and, thus, turbulence and mixing. The instability is suppressed if the wavelength of the perturbation ($\lambda$) is smaller than $\sim 10 d$ \citep{Chandrasekhar_1961}. 

We assume that the instability will be fully suppressed when the viscous timescale, $\tau_{\nu}=d^2/\nu$, at which the velocity gradient is smoothed out a distance $d = \lambda / 10$ is shorter than the Kelvin-Helmholtz timescale,\footnote{This is the equation of the KH timescale at the incompressible limit (see Sect. \ref{sec:crit_visc_mach}).} 
\begin{equation}
    \tau_{\rm KH} = \frac{\lambda}{\Delta v_{\rm shear}} \frac{(\rho_{\rm hot} + \rho_{\rm cold})}{(\rho_{\rm hot}\,\rho_{\rm cold})^{1/2}} \, .
    \label{eqn:tau_KH}
\end{equation}
Here, $\rho_{\rm cold}$ and $\rho_{\rm hot}$ are the densities of the cold and hot medium respectively, $\Delta v_{\rm shear}$ is the velocity difference between the two media, and $\nu$ is the kinematic viscosity, which depends on the shear viscosity coefficient via 
\begin{equation}
    \nu = \eta / \rho \, .
    \label{eqn:kinematic_coeff}
\end{equation}
The instability becomes stable and, therefore, fully suppressed when
\begin{equation}
    \tau_{\nu} < \tau_{\rm KH},
\end{equation}
which can be rewritten to
\begin{equation}
    \nu > \nu_{\rm Crit} = \frac{\lambda \, \Delta v_{\rm shear}}{100} \frac{(\rho_{\rm hot}\,\rho_{\rm cold})^{1/2}}{(\rho_{\rm hot} + \rho_{\rm cold})} \, .
    \label{eqn:nu_crit}
\end{equation}

\subsection{Radiative mixing layers} \label{sec:theory_cooling}

When the KHI develops, it quickly triggers turbulence, mixing the cold and hot medium into intermediate temperature gas, entering the regime where the cooling curve peaks \citep{Begelman_1990,Fielding_2020,Tan_2021}. To characterize the evolution it is convenient to introduce the Damköhler number \citep[Da;][]{Damkohler_1940}, a dimensionless number that accounts for the cooling strength:
\begin{equation}
    {\rm Da} = \frac{t_{\rm turb}}{t_{\rm cool} (T)} = \frac{\lambda}{u' \, t_{\rm cool} (T)} \, ,
    \label{eqn:Da_theory}
\end{equation} 
where $t_{\rm turb}$ is the eddy turnover time, $t_{\rm cool}$ the cooling time, and $u'$ the turbulent velocity.

When $t_{\rm turb} < t_{\rm cool}$ (${\rm Da} < 1$), the effect of cooling is weak compared to turbulence, and the gas entropy is dominated by mixing. This leads to a characteristic mass flux $\propto t_{\rm cool}^{-1/2}$ \citep{Ji_2019,Tan_2021}. However, when $t_{\rm cool} < t_{\rm turb}$ (${\rm Da} > 1$), cooling is stronger than turbulence and, as soon as the gas is mixed, it is cooled down, leading to a multiphase medium where the gas entropy is dominated by cooling \citep[see, e.g.,][]{Kuo_2012, Tan_2021}. Specifically, in this strong cooling regime, the total luminosity of the mixing layer scales as $\propto t_{\rm cool}^{-1/4}$ \citep{Gronke_2019,Fielding_2020,Tan_2021}.

In summary, the emission due to cooling follows a power-law relation, which depends on the cooling strength (Damköhler number). The surface brightness $Q$ depends on ${\rm Da}$ via 
\begin{equation}
Q \propto
\begin{cases}
    u'^{\,1/2} \lambda^{1/2} \propto u' {\rm Da}^{1/2}, & \text{if } \mathrm{Da} < 1 \text{ (weak cooling);} \\
    u'^{\,3/4} \lambda^{1/4} \propto u' {\rm Da}^{1/4}, & \text{if } \mathrm{Da} > 1 \text{ (strong cooling)} \, .
\end{cases}
\label{eqn:Da_Q_u}
\end{equation}

\section{Numerical setup} \label{sec:numerical_setup}
We performed our simulations using the code \textsc{Athena} \citep{Stone_2008}. We used the HLLC Riemann solver, a second-order reconstruction with slope limiters in the primitive variables, the van Leer unsplit integrator \citep{Gardiner_2008}, Cartesian geometry, and an adiabatic equation of state with $\gamma=5/3$. In all our simulations, we kept the shear viscosity coefficient ($\eta$) constant to simplify the analysis and understand the effect of viscosity in the simplest way\footnote{For a discussion on the effect of temperature-dependent viscosities, see Sect. \ref{sec:hot_vs_cold_viscosity}.}.

\subsection{Adiabatic 2D simulations}

To better understand fluid dynamics and the effects of viscosity while reducing computational costs, we used two different 2D setups without cooling. These provided a more controlled environment compared to a chaotic 3D system, while allowing us to study the KHI in detail. The first setups explore how critical viscosity depends on overdensity and Mach number by varying viscosity across different overdensity and Mach number values. We describe each 2D setup in detail below.

\subsubsection{The planar slab} \label{sec:planar_slab}

To understand how the critical viscosity at which the KHI is suppressed changes as a function of the overdensity, we used a modified version of the numerical setup suggested by \citet{Lecoanet_2015}, which provides a benchmark for the growth of the KHI in grid codes. We created a 2D domain with a resolution of $320\times320$ in the $\hat{x}$ and $\hat{y}$ directions, respectively, with periodic boundary conditions with dimensions $L_x = 256$ and $L_y = 256$. The density and velocity profiles satisfy
\begin{equation}
    \rho = \rho_{\mathrm{hot}} + \frac{\rho_{\mathrm{cold}}-\rho_{\mathrm{hot}}}{2}\left[\tanh\left(\frac{y+y_{\mathrm{Int}}}{a}\right) - \tanh\left(\frac{y-y_{\mathrm{Int}}}{a}\right)\right] \, ,
\end{equation}
\begin{equation}
    v_x = -v_{\mathrm{shear}} \times \left[\tanh\left(\frac{y+y_{\mathrm{Int}}}{a}\right) - \tanh\left(\frac{y-y_{\mathrm{Int}}}{a}\right) - 1 \right] \, .
\end{equation}
Here, $\rho_{\mathrm{hot}}$ is a fixed value for all our simulations and we end up modifying $\rho_{\mathrm{cold}}$ as we change the value of overdensity. Then, $y_{\mathrm{Int}} = 64$ indicates the position of the interface between the two fluids and $v_{\mathrm{shear}}$ is the shear velocity of each fluid; namely, the relative motion can be expressed as $\mathcal{M}\equiv 2v_{\mathrm{shear}}/c_{\mathrm{c, hot}}$.

In this setup, we kept $\mathcal{M} = 0.34$ constant, as well as $T_{\mathrm{cold}} = 10^4$ K, while $T_{\mathrm{hot}}$ was given by the overdensity to ensure pressure equilibrium in the whole domain. Overall, the value of $a = \lambda/10$ ensures smooth initial conditions, fundamental for the growth of the desired instability mode without spurious secondary instabilities in grid codes; $\lambda = 128$ indicates the wavelength of the perturbation used to trigger the instability. We used the same initial perturbation as in \citet{Marin-Gilabert_2022}, a modified version of the one introduced by \citet{Read_2010}, expressed as
\begin{multline}
    v_y = -\delta v_y \left[ \sin \left( \frac{2\pi (x+\lambda/2)}{\lambda}\right) \exp \left(-\left(\frac{y-y_{\mathrm{Int}}}{\sigma}\right)^2\right) + \right. \\
    \left. + \sin \left( \frac{2\pi x}{\lambda}\right) \exp \left(-\left(\frac{y+y_{\mathrm{Int}}}{\sigma}\right)^2\right) \right] \, ,
    \label{eqn:init_pert_2D}
\end{multline}
where $\delta v_y = v_{\mathrm{shear}}/10$ is the initial amplitude of the perturbation and $\sigma = 0.2\lambda$ is a scaling parameter to control the width of the perturbation layer.

To study how the critical viscosity changes with the overdensity of the system, we used different values of $\chi\equiv \rho_{\mathrm{cold}}/\rho_{\rm hot}$ from 1 to 100 and tested different values of viscosity for each overdensity. For reference, the Spitzer viscosity of a fluid with $T = 10^6$~K is $\eta_{\rm Sp} \simeq 0.11$g(cm s)$^{-1}$, which corresponds to $\eta_{\rm Sp} \simeq 1809.28$ in our unit system.

\subsubsection{The planar sheet} \label{sec:planar_sheet}

When increasing the Mach number, body modes caused by pressure waves moving across the domain between the stream boundaries appear in the planar slab setup \citep{Mandelker_2019}. To avoid this, we modified our initial conditions to study the dependence of the critical viscosity as a function of the Mach number. We used a similar setup as before, but with just one interface between the two fluids at $y_{\mathrm{Int}} = 0$ and open boundaries in the $\hat{y}$ direction. The density and velocity profiles are expressed as\begin{equation}
    \rho = \frac{1}{2} \times \left[\rho_{\mathrm{cold}} \left(1-\tanh\left(\frac{y}{a}\right)\right) + \rho_{\mathrm{hot}} \left(1+\tanh\left(\frac{y}{a}\right)\right) \right] \, ,
    \label{eqn:dens_gradient}
\end{equation}
\begin{equation}
    v_x = v_{\mathrm{shear}} \times \tanh\left(\frac{y}{a}\right) \, ,
\end{equation}
and the initial perturbation is
\begin{equation}
    v_y = -\delta v_y \sin \left( \frac{2\pi (x+\lambda/2)}{\lambda}\right) \exp \left(-\left(\frac{y}{\sigma}\right)^2\right) \, .
\end{equation}

With this setup, we want to study the dependence of the critical viscosity on the Mach number. Therefore, we kept  $\chi=10$ constant and we modified $v_{\mathrm{shear}}$  to obtain  Mach values ranging from 0.01 to 1.8. As with the previous setup, we tested different values of viscosity for each Mach number to identify the critical viscosity in each case.

\subsection{3D simulations: Turbulent radiative mixing layers} \label{sec:3D_setup}

Once we obtained the value for the critical viscosity in a system with an overdensity of 100, we made use of this value to study how different values of viscosity affect a TRML system. For a proper development of turbulence, a 3D setup is needed; therefore, we used a modified version of the setup suggested in \citet{Ji_2019}, where the density profile is the same as in Eq. \ref{eqn:dens_gradient}, but the velocity profile is
\begin{equation}
    v_x = \frac{v_{\mathrm{shear}}}{2} \times \left[1 - \tanh\left(-\frac{y}{a}\right) \right] \, ,
\end{equation}
while the initial perturbation is
\begin{equation}
    v_y = \delta v_y \sin \left( \frac{2\pi x}{\lambda}\right) \sin \left( \frac{2\pi z}{\lambda}\right) \exp \left(-\left(\frac{y}{a}\right)^2\right) \, .
\end{equation}

In this case, $v_{\mathrm{shear}}$ corresponds to the velocity difference between the media, where the lower (colder) medium is stationary and the upper one moves at $v_{\mathrm{shear}}$. We used a Mach number of $\mathcal{M} = 0.5$ and $\delta v_y = v_{\mathrm{shear}} / 100$. In this case our domain is ten times larger in the $\hat{y}$ direction, whereas in the $\hat{x}$ and $\hat{z}$ direction is equal to the wavelength of the perturbation $L_y = 10L_x = 10L_z = 10\lambda$. We kept the hot gas density and the cold gas temperature constant, $\rho_{\mathrm{hot}}$ and $T_{\mathrm{cold}} = 10^4$~K, respectively. In this setup, we use a fixed overdensity $\chi = 100$ (i.e., $T_{\mathrm{hot}} = 10^6$~K). The resolution is the same as in \citet{Das_2023}, consisting of $64 \times 640 \times 64$ in the $\hat{x}$, $\hat{y}$, and $\hat{z}$ directions respectively, with outflow boundaries in the $\hat{y}$ direction and periodic boundaries in $\hat{x}$ and $\hat{z}$.

We used the Townsend radiative cooling algorithm \citet[as, e.g., in \citealp{Gronke_2022}]{Townsend_2009}, with a solar metallicity cooling curve \citep{Sutherland_1993} and a $T_{\mathrm{floor}} = 10^4$~K. In contrast to \citet{Tan_2021}, we did not modify the cooling strength. Instead, we modified the wavelength to have different values of Damköhler number. Since $\mathrm{Da} = t_{\mathrm{turb}} / t_{\mathrm{cool}}$, and $t_{\mathrm{turb}} \propto \lambda$ (see Eq. \ref{eqn:Da_theory}), by varying the wavelength (and also the size of the box), we got different values of Da. We ran a total 12 simulations with different Da for each amount of viscosity, from $\mathrm{Da_{mix}} \simeq 2.3\times10^{-3}$ to $\mathrm{Da_{mix}} \simeq 1.1\times10^{3}$, where the temperature of the intermediate mixed gas chosen to compute the cooling time is $T_{\mathrm{mix}} = 2\times10^5$~K (to be consistent with \citealp{Tan_2021}). The values of Da were calculated assuming that $t_{\mathrm{turb}} \sim \tau_{\rm KH}$. We know that in the presence of cooling, the box is filled with cold gas, causing the mixing layer leaves the domain, especially in the strong cooling regime. To prevent this, we can calculate the velocity at which the box is being filled with cold gas,  adding a velocity in the opposite direction to the whole box, following \citet{Das_2023}.

\section{Results}

\subsection{Critical viscosity versus overdensity} \label{sec:crit_visc_overdens}

We provide an analytical estimate for the critical viscosity in a KHI setup in Sect. \ref{sec:theoretical_estimate} (see Eq.~\ref{eqn:nu_crit}, based on the growth time of the KHI and the viscous time. Numerically, this critical viscosity is computed by measuring the growth or decay of the instability within 1$\tau_{\rm KH}$ (see Appendix \ref{app:growth_KHI} for details). 

The analytical estimate presented in Sect. \ref{sec:theoretical_estimate} sets a criterion for the kinematic viscosity ($\nu$). However, in our simulations, we used a constant dynamic viscosity ($\eta$) for both fluids. In a multi-fluid system with different densities but the same $\eta$, each fluid will exhibit its own $\nu$. Therefore, to be able to compute the critical $\eta$ for our setup, it is necessary to define an appropriate average of these $\nu$ to find a general $\bar{\nu}$ of the whole system. Thus, for a system characterized by a single dynamic viscosity ($\eta$), two densities ($\rho_{\rm h}$ and $\rho_{\rm c}$), and consequently two different kinematic viscosities ($\nu_{\rm h}$ and $\nu_{\rm c}$), we define an averaged kinematic viscosity using the harmonic mean, namely,
\begin{equation}
    \nu_{\rm h} = \frac{\eta}{\rho_{\rm h}} \, ,\hspace{1cm} \, \nu_{\rm c} = \frac{\eta}{\rho_{\rm c}}\, ,
\end{equation}
\begin{equation}
    \bar{\nu} = \frac{\sum_i \eta_i}{\sum_i \rho_i} = \frac{2 \eta}{\rho_{\rm h} + \rho_{\rm c}} = \frac{2 \eta}{\frac{\eta}{\nu_{\rm h}} + \frac{\eta}{\nu_{\rm c}}} = \frac{2}{\frac{1}{\nu_{\rm h}} + \frac{1}{\nu_{\rm c}}} \, ,
\end{equation}
which is equivalent to a density weighted mean,
\begin{equation}
    \bar{\nu} = \frac{\nu_{\rm h} \rho_{\rm h} + \nu_{\rm c} \rho_{\rm c}}{\rho_{\rm h} + \rho_{\rm c}} = \frac{2 \eta}{\rho_{\rm h} + \rho_{\rm c}} \rightarrow \eta = \frac{\bar{\nu} \, (\rho_{\rm h} + \rho_{\rm c})}{2}\, .
\end{equation}
Taking the expression of critical kinematic viscosity (see Eq. \ref{eqn:nu_crit}), and assuming that $\bar{\nu}$ corresponds to the averaged value of the two systems, we can express the value of the critical dynamic viscosity of our system as
\begin{equation}
    \eta_{\rm Crit} = \frac{\lambda \, \rho_{\rm h} \, \mathcal{M}_{\rm h} c_{\rm s, cold} \, \chi}{200} \, ,
    \label{eqn:critical_eta}
\end{equation}
where the wavelength ($\lambda$), the density and Mach number of the hot gas ($\rho_{\rm h},\, \mathcal{M}_{\rm h}$) and the sound speed of the cold gas ($c_{\rm s, cold}$) are kept constant.

Previous studies have tested the critical viscosity estimate for different overdensities, but the results deviated from the predictions at high overdensities since they did not consider the difference in momentum \citep[e.g.,][]{Esch_1957, Roediger_2013}. For high overdensities each fluid carries a very different momentum (inertia) which viscosity needs to compensate for. By using an harmonic (weighted) mean, we are taking this effect into account.

\begin{figure}
   \centering
   \includegraphics[width=\hsize]{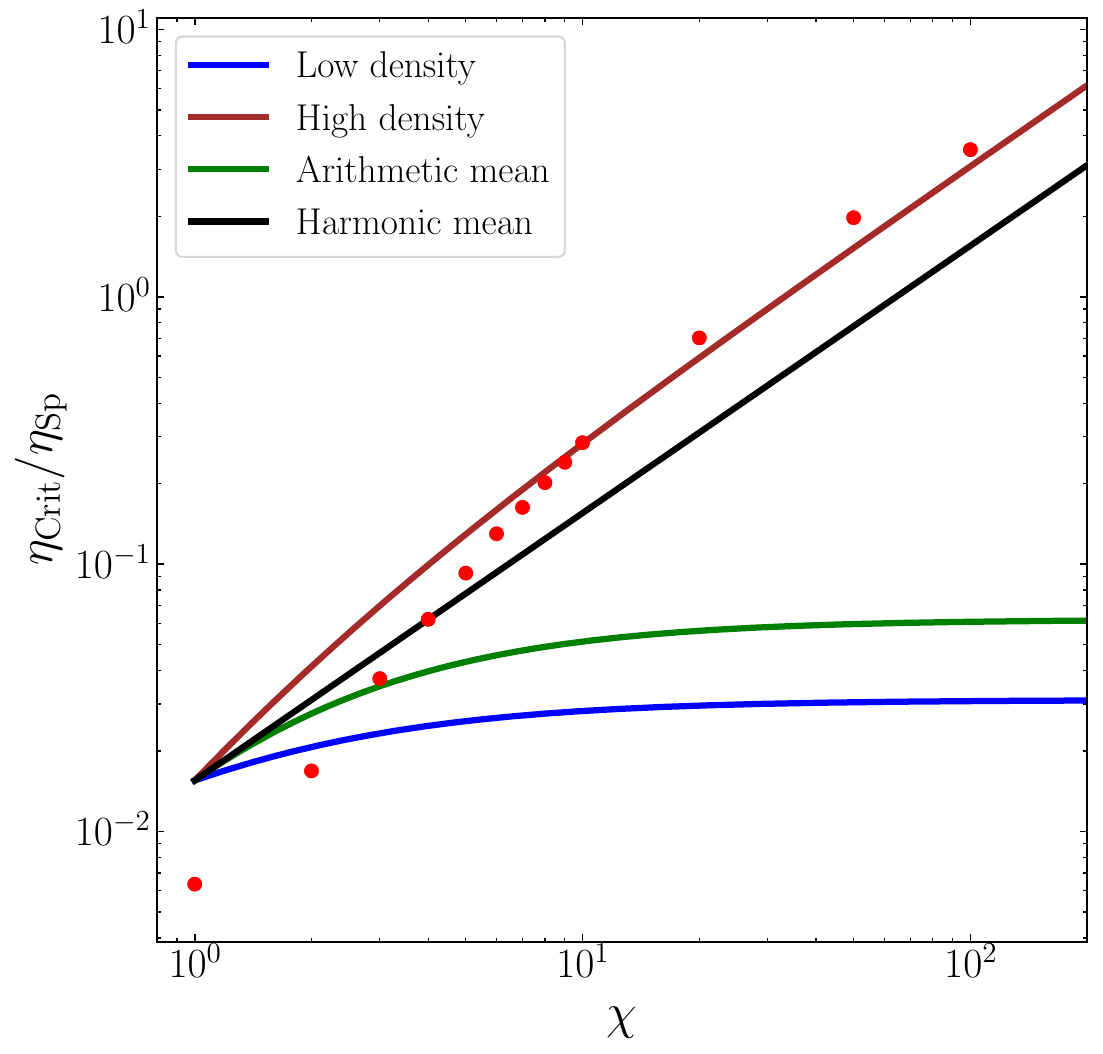}
      \caption{Dependence of $\eta_{\mathrm{Crit}}$ on the overdensity, $\chi$. Red dots show the numerical results. The blue line shows the theoretical expectation assuming that only the low density is relevant for kinematic viscosity. The red line shows that only the high density is relevant. The green line assumes an arithmetic mean of kinematic viscosity between the fluids. The black line shows a harmonic (weighted) mean. Note that we normalized the values of viscosity to the Spitzer value at $10^6\,$K.}
    \label{fig:crit_visc_chi}
\end{figure}

To test the dependence of a critical viscosity on the overdensity $\chi$, we ran different simulations changing the value of $\chi$ and measure the critical viscosity in each case, using the 2D setup described in Sect.  \ref{sec:planar_slab}. Since we included a unique dynamic viscosity in our setup, we quantified the critical dynamic viscosity (Eq. \ref{eqn:critical_eta}). Figure~\ref{fig:crit_visc_chi} shows the numerical results (red dots) normalized to the Spitzer value of viscosity at $T=10^6$~K, together with different methods to average the kinematic viscosity besides the harmonic mean (black line): considering only the low density fluid $\bar{\nu} = \eta/\rho_{\rm h}$ (blue line); considering only the high density fluid $\bar{\nu} = \eta/\rho_{\rm c}$ (red line); and finding an arithmetic mean $\bar{\nu} = (\nu_{\rm h} + \nu_{\rm c}) / 2$ (green line).

The cold medium becomes denser when increasing the density contrast. This has two effects: \textit{(i)} the cold medium has a much lower kinematic viscosity, $\nu_{\rm c} = \eta/\rho_{\rm c}$. \textit{(ii)} The cold medium carries much more momentum than the hot medium. For the high overdensity flows $\chi \gg 1$ we consider here, it is a good approximation to assume that the viscosity of the cold gas is negligible (as we show explicitly in Sect. \ref{sec:hot_vs_cold_viscosity}, the hot gas viscosity is what matters), but that it contains almost all the momentum. Thus, the critical viscosity increases with overdensity, $\eta_{\rm crit} \propto \chi$ (see Eq.~\ref{eqn:critical_eta} and Fig.~\ref{fig:crit_visc_chi}) because the total momentum in the flow is higher and stronger friction is required to have an effect.  

From the values of critical viscosity obtained and using the harmonic mean, we can compute their corresponding Reynolds number as
\begin{equation}
    {\rm Re_{Crit}} = \frac{\lambda \, v_{\rm shear}}{\bar{\nu}_{\rm Crit}} = \frac{\lambda \, v_{\rm shear} (\rho_{\rm h} + \rho_{\rm c})}{2\eta_{\rm Crit}} = 100 \frac{1+\chi}{\chi^{1/2}},
\end{equation}
namely, ${\rm Re}_{\rm Crit}\approx 100 \chi^{1/2}$ for large $\chi$.

\subsection{Critical viscosity versus Mach number} \label{sec:crit_visc_mach}

According to our analytical estimate, the critical viscosity depends linearly on the Mach number for a given overdensity (see Eq. \ref{eqn:critical_eta}). To test this relation, we ran different simulations varying $\mathcal{M}_{\rm h}$, while keeping $\chi = 10$ fixed and we measured the critical viscosity in each case, using the setup described in Sect.  \ref{sec:planar_sheet}. However, the theoretical analysis was made assuming incompressible fluids, which is a good approximation for low Mach number, but breaks down at high Mach numbers, where compressibility effects become significant. 

When compressibility is taken into account, the classical dispersion relation of the KHI derived from linear analysis is no longer valid \citep[][]{Landau_1944, Landau_1987}. Instead, the dispersion relation for compressive fluids in a planar sheet is
\begin{equation}
    \left[ \chi (\varpi - 1)^2 - \varpi^2 \right] \cdot \left[\chi (\varpi - 1)^2 (\mathcal{M}_{\rm h}^2 \varpi^2 - 1) - \varpi^2 \right] = 0 \, ,
    \label{eqn:mandelker_im_w}
\end{equation}
where $\varpi \equiv \omega/k\Delta v_{\rm shear}$ (see \citealp{Mandelker_2016} for details).

\begin{figure}
   \centering
   \includegraphics[width=\hsize]{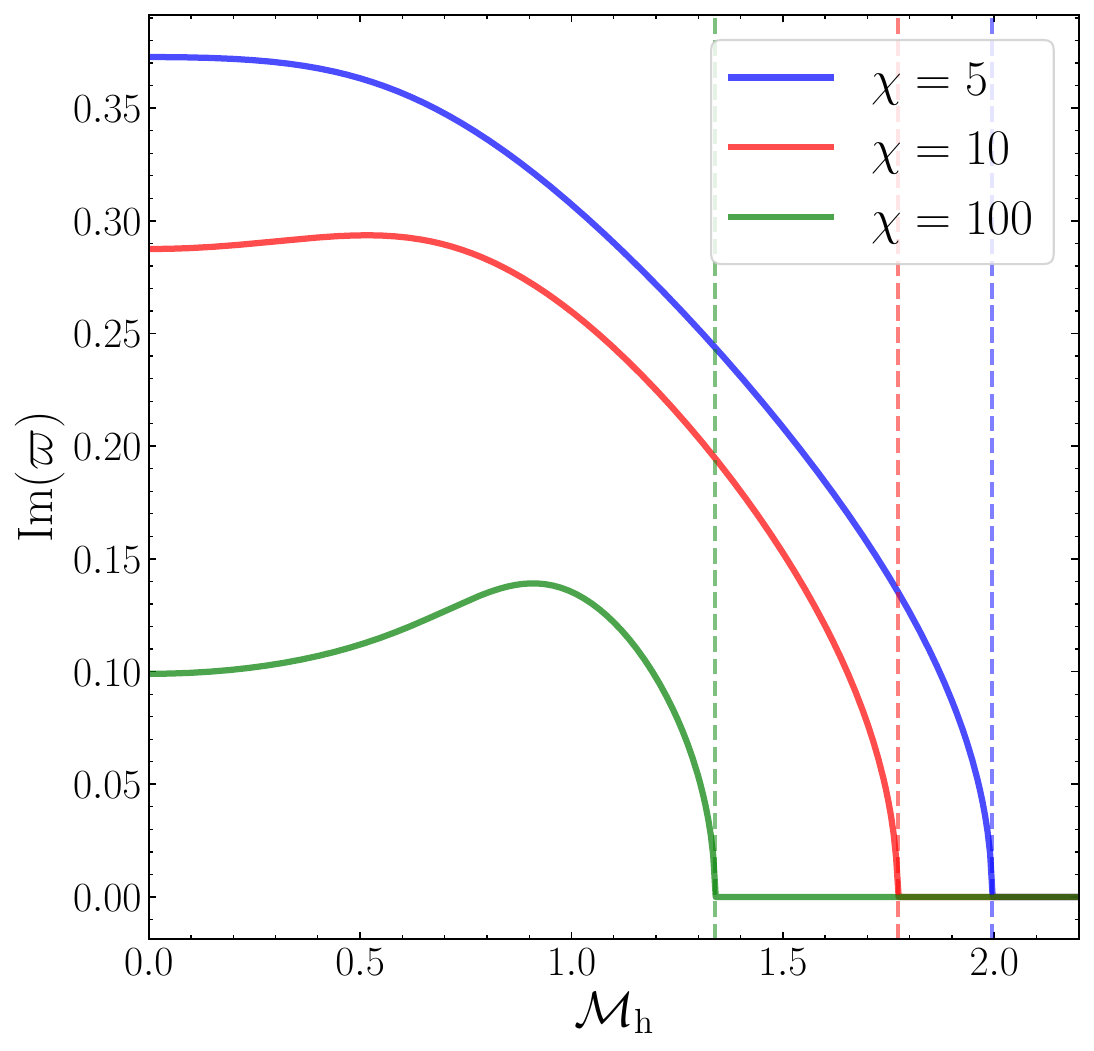}
      \caption{Numerical solution for the values of Im($\varpi$) of Eq. \ref{eqn:mandelker_im_w} as a function of $\mathcal{M}_{\mathrm{h}}$ for $\chi = 5, 10, 100$. The dashed lines show the $\mathcal{M}_{\mathrm{h}}$ above which the KHI is stable.}
      \label{fig:w_mandelker}
\end{figure}

The two roots of the quadratic part of the equation are always real; therefore, they do not describe the growth of the KHI. Two of the four roots in the quartic factor are also real, while two are complex conjugates. We took the imaginary part of both complex conjugates, which describe an exponential growth (positive sign) or decay (negative sign) of the instability. The positive imaginary numerical solutions of Eq. \ref{eqn:mandelker_im_w} for $\chi = 5, 10, 100$ are shown in Fig.~\ref{fig:w_mandelker}.

In the case of $\chi = 10$, the instability grows (positive Im($\varpi$)) for $\mathcal{M}_{\rm h} \lesssim 1.77$. Above this critical value of Mach number ($\mathcal{M}_{\rm Crit}$), the solution becomes real and the KHI is stable (vertical dashed lines). The general solution for $\mathcal{M}_{\rm Crit}$  \citep[][]{Mandelker_2016} is given by
\begin{equation}
    \mathcal{M}_{\rm Crit} = \left(1 + \chi^{-1/3}\right)^{3/2}.
\end{equation}
The value of $\mathcal{M}_{\rm Crit}$ for $\chi = 10$ is computed numerically in Fig.~\ref{fig:growth_mach_0eta} for nonviscous fluids using the method described in Appendix \ref{app:growth_KHI}, finding a $\mathcal{M}_{\rm Crit} = 1.65 \pm 0.5$. 
\begin{figure}
   \centering
   \includegraphics[width=\hsize]{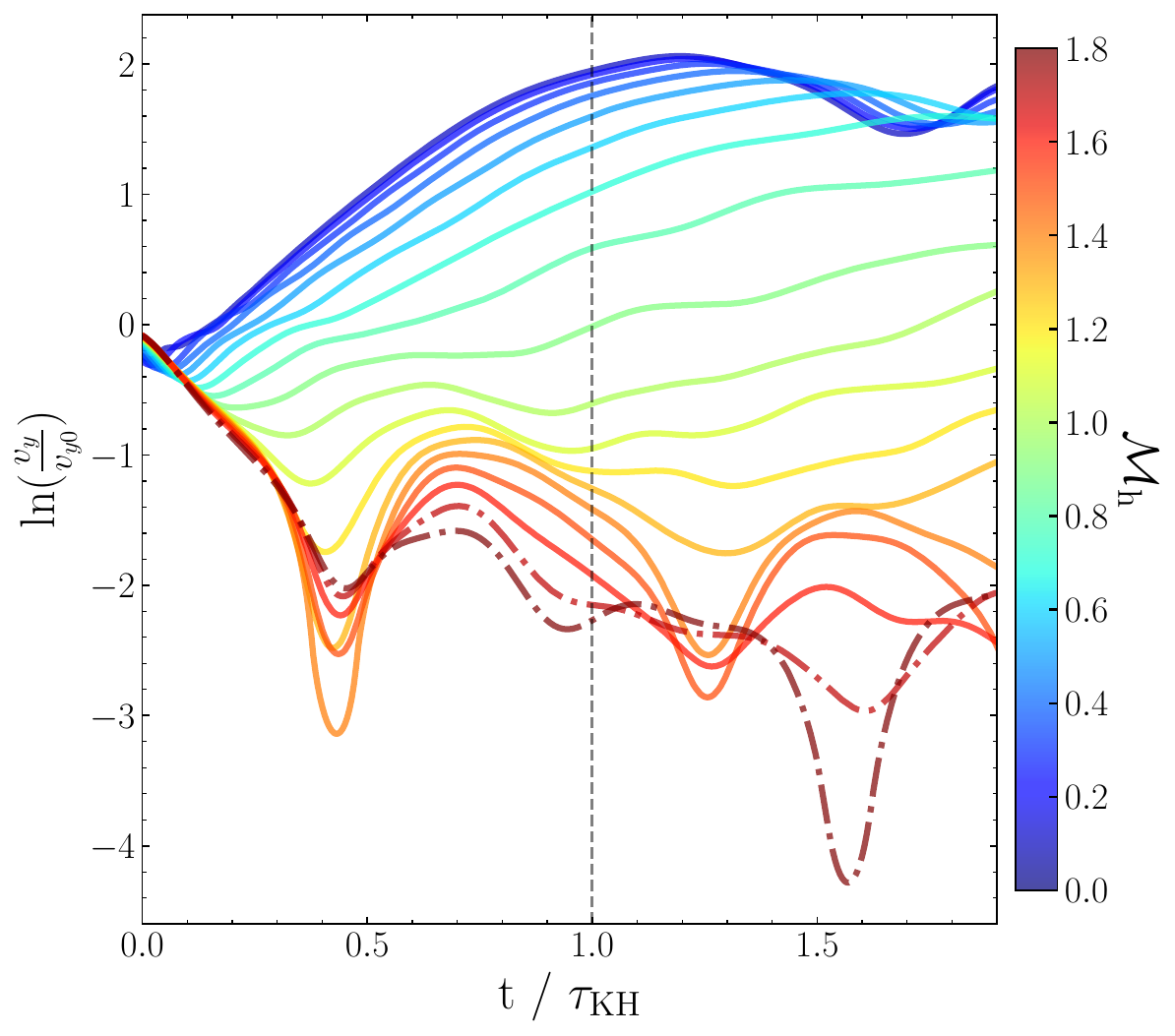}
      \caption{Growth of the KHI with time for $\chi = 10$, color-coded by $\mathcal{M}_{\mathrm{h}}$. The solid lines indicate the simulations in which the KHI is still unstable and the dash-dotted lines where the instability is suppressed (values above $\mathcal{M}_{\mathrm{Crit}}$), finding a $\mathcal{M}_{\rm Crit} = 1.65 \pm 0.5$.}
      \label{fig:growth_mach_0eta}
\end{figure}

For $\mathcal{M}_{\rm h} > \mathcal{M}_{\rm Crit}$, the instability is suppressed and the growth rate decays. An intuitive explanation for this is that when the instability starts growing, areas of high and low-pressure are formed around the perturbation at a rate of approximately the sound-crossing time. These areas feed the instability, leading to its growth. At large mach number, the shear motion of the fluids is faster than the sound-speed, preventing these areas of high and low pressure from forming and leading to the flattening and stabilization of the wave.

For $\mathcal{M}_{\rm h} > \mathcal{M}_{\rm Crit}$, the instability is fully suppressed by the compressive nature of the fluids and the viscosity needed to keep the system stable is zero. Therefore, there is a transition from the incompressible limit where the suppression is dominated by viscosity to the compressible limit set by $\mathcal{M}_{\rm Crit}$. An analytical solution for $\eta_{\rm Crit}$ as a function of $\mathcal{M}_{\rm h}$ can be estimated in a similar way, as in Sect.  \ref{sec:theoretical_estimate}.

In the incompressible limit ($\mathcal{M}_{\rm h} \rightarrow0$), the positive imaginary solution of Eq. \ref{eqn:mandelker_im_w} yields
\begin{equation}
    {\rm Im}(\varpi) = \frac{\sqrt{\chi}}{(\chi + 1)} \, , \hspace{1cm}
    {\rm Im}(\omega) = k \Delta v \, \frac{\sqrt{\chi}}{(\chi + 1)} \, ,
\end{equation}
recovering the KH timescale (Eq. \ref{eqn:tau_KH}), which, in its general form, is defined as 
\begin{equation}
    \tau_{\rm KH} = \frac{1}{k \Delta v_{\rm shear} \, {\rm Im(\varpi)}} \, .
\end{equation}
By using the numerical solution for ${\rm Im}(\varpi)$ (shown in Fig.~\ref{fig:w_mandelker})  and evaluating $\tau_{\rm KH}>\tau_{\nu}$ (i.e., the same analysis as in Sect.  \ref{sec:theoretical_estimate} for the incompressible case), we obtained an analytical solution for a critical viscosity in the compressible case,
\begin{equation}
    \eta_{\rm Crit} = \frac{\lambda \, \rho_{\rm h} \, \mathcal{M}_{\rm h} c_{\rm s, cold} \sqrt{\chi} \, (\chi + 1) \, {\rm Im(\varpi)}}{200} \, .
    \label{eqn:crit_compressible_eta}
\end{equation}

\begin{figure}
   \centering
   \includegraphics[width=\hsize]{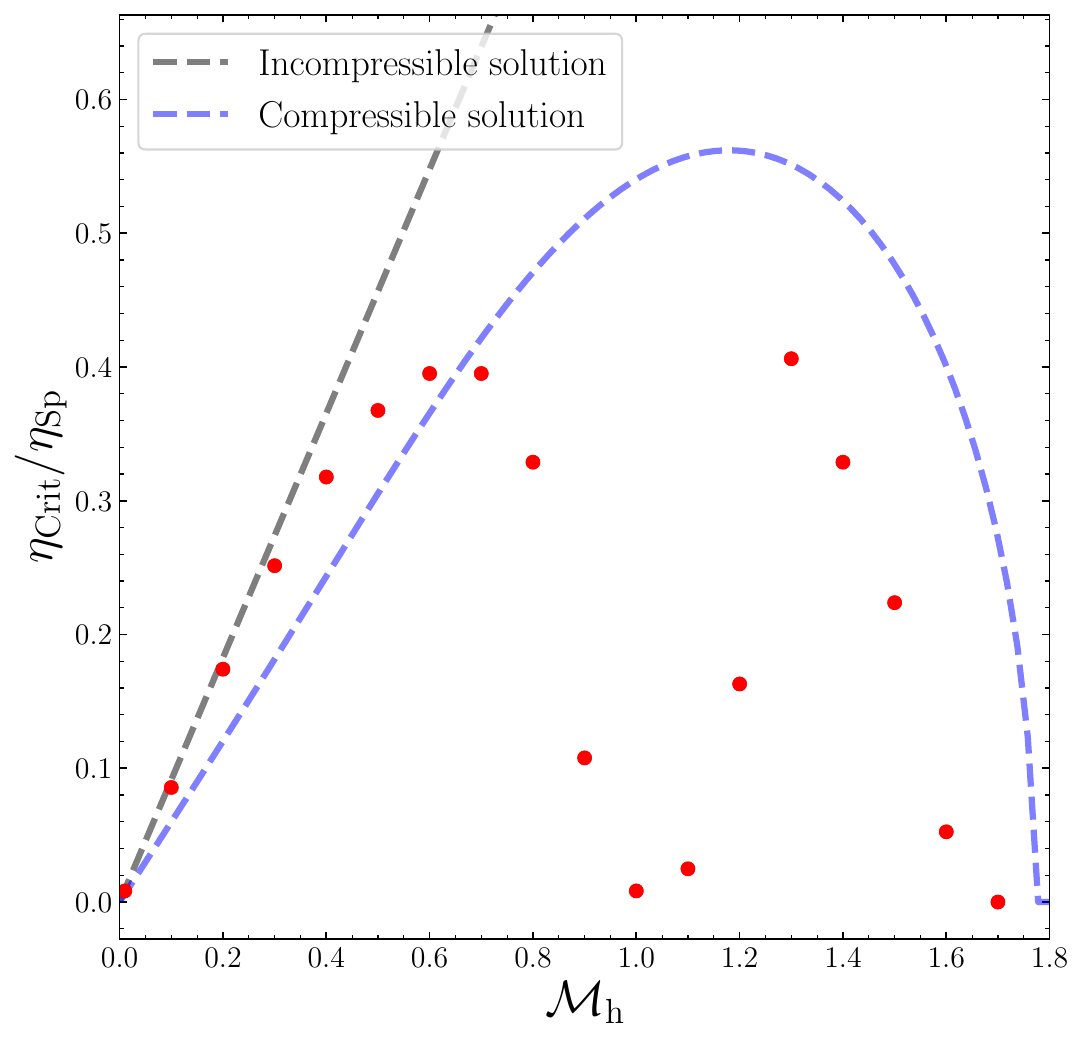}
      \caption{Dependence of $\eta_{\mathrm{Crit}}$ on the Mach number, $\mathcal{M}_{\mathrm{h}}$. The numerical results (red dots) deviate from the analytical incompressible $\eta_{\mathrm{Crit}}(\mathcal{M}_{\mathrm{h}})$ (Eq.~\ref{eqn:critical_eta}; dashed black line). However, this is corrected by taking into account the suppression due to high $\mathcal{M}_{\mathrm{h}}$ (Eq.~\ref{eqn:crit_compressible_eta}, dashed blue line). Note that we normalized the viscosities to the Spitzer value at $10^6\,$K.}
      \label{fig:crit_visc_mach}
\end{figure}

In Fig.~\ref{fig:crit_visc_mach} we show  the analytical estimate of $\eta_{\rm Crit}$, normalized to the Spitzer value at $T=10^6$~K, as a function of $\mathcal{M}_{\rm h}$ for incompressible fluids (dashed black line) and for compressible fluids (dashed blue line). The red data points are the numerical values of $\eta_{\rm Crit}$ obtained from the simulations. Although the fit is not perfect due to the large assumptions made in the estimate, the initial growth at low $\mathcal{M}_{\rm h}$, and the decay when $\mathcal{M}_{\rm h} \sim \mathcal{M}_{\rm Crit}$ match the theoretical estimate. However, there is a drop at $\mathcal{M}_{\rm h} \sim 1$, indicating that the KHI becomes stable at transonic speeds. Although the origin of this drop is not clear, one possibility is that higher order modes of the KHI have different values of $\eta_{\rm Crit}$. We will explore this purely adiabatic suppression in a future work.

\subsection{Turbulent radiative mixing layers} \label{sec:TRMLs}

To study how viscosity affects the TRMLs, we made use of the value of critical viscosity and Reynolds number obtained in Sect. \ref{sec:crit_visc_overdens} for $\chi = 100$: $\eta_{\rm Crit} = 6425\pm25$  and Re$_{\mathrm{Crit}} = 248.6 \pm 2.4$ respectively. In terms of Spitzer viscosity, the critical viscosity measured corresponds to $\eta_{\rm Crit} \simeq 3.55 \eta_{\rm Sp}$ at a temperature of $T = 10^6$~K. We express the amount of viscosity in our setup as a function of the critical viscosity needed to suppress the instability.
Using the 3D setup described in Sect. \ref{sec:3D_setup}, we ran a simulation with no viscosity for each Damköhler number Da (labeled ``No visc''), with 5\% of the critical viscosity (``$0.05\eta_{\mathrm{Crit}}$''), 10\% of critical viscosity (``$0.1\eta_{\mathrm{Crit}}$'') and 50\% of critical viscosity (``$0.5\eta_{\mathrm{Crit}}$''). Although the definition of critical viscosity considers only the growth of the KHI within $1\tau_{KH}$, in this step, we ran the simulations for longer times for the proper development of turbulence. The set of simulations with different viscosities, as well as the value of the Reynolds number in each case are shown in Table~\ref{tab:TRMLs_runs}.
\begin{table}
    \centering
    \caption{Simulations of viscous, radiative mixing layers.}
    \renewcommand\tabcolsep{2.mm}
    \begin{tabular}{c c c c}
        \hline
         Label & Fraction of $\eta_{\rm Crit}$ & $\rm Re$ & Spitzer fraction \\
        \hline 
        No visc & - & - & - \\
         
        $0.05\eta_{\mathrm{Crit}}$ & 5\% of $\eta_{\mathrm{Crit}}$ & $4972\pm48$ & $0.18 \eta_{\rm Sp}$ \\
        
        $0.1\eta_{\mathrm{Crit}}$ & 10\% of $\eta_{\mathrm{Crit}}$ & $2486\pm24$ & $0.36 \eta_{\rm Sp}$ \\
        
        $0.5\eta_{\mathrm{Crit}}$ & 50\% of $\eta_{\mathrm{Crit}}$ & $497.2\pm4.8$ & $1.78 \eta_{\rm Sp}$ \\
        \hline
    \end{tabular}
    \label{tab:TRMLs_runs}
\end{table}

It is important to note that in the "No visc" run, although we did not include any explicit physical viscosity, there is still numerical viscosity given by resolution. Since our resolution consists of $64 \times 64$ in the $\hat{x}\hat{z}$ plane and the length of the box is $L_x = L_z = \lambda$, the cell length is $\Delta x = \lambda / 64$, which sets the Kolmogorov scale of our "No visc" runs. The runs with explicit viscosity set the Kolmogorov scale at larger values, given by percentage of critical viscosity taken in each case. 
Since we have not included explicit thermal conduction, the Field length of thermal conduction is given by $\delta = \Delta x = \lambda / 64$. This allows us to explore the Karlovitz number (see Sect. \ref{sec:introduction}) parameter space from $\mathrm{Ka} = 1$ ("No visc" run) toward smaller values of Ka.

\subsubsection{Front morphology}

\begin{figure}
   \centering
   \includegraphics[width=\hsize]{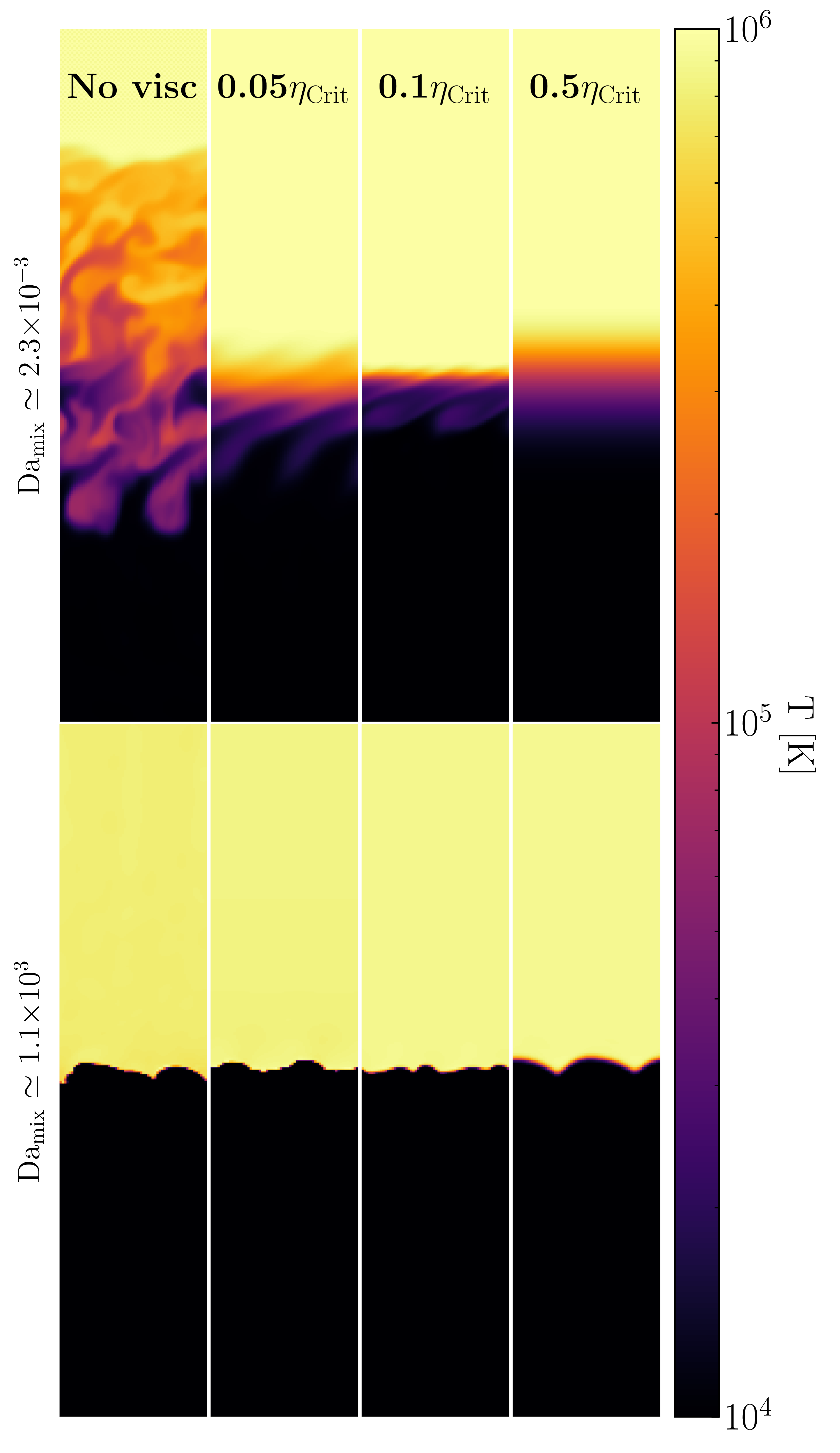}
      \caption{Temperature slices for the different TRML simulations. \textit{From left to right}: Non-viscous case, 5\% of critical viscosity, 10\% of critical viscosity and 50\% of critical viscosity. \textit{Top}: Very weak cooling with $\mathrm{Da_{mix}} \simeq 2.3\times10^{-3}$. \textit{Bottom}: Very strong cooling with $\mathrm{Da_{mix}} \simeq 1.1\times10^{3}$.}
      \label{fig:temp_slices}
\end{figure}

In Fig.~\ref{fig:temp_slices} we show how viscosity affects mixing in the different runs for a low Da (weak cooling, top panels) and high Da (strong cooling, bottom panels). Each panel shows a temperature slice through the box, centered at the interface between the two fluids at $t = 2.5\tau_{\mathrm{KH}}$. In the weak cooling regime (low Da, top panels), turbulence has developed in the run without explicit viscosity whereas it is more suppressed the more viscous the media are (left to right).

It is important to note that the computation of critical viscosity is done during $t < \tau_{\mathrm{KH}}$, but naturally viscosity is affecting the fluid movement for the entire simulation. This implies that, although during $t < \tau_{\mathrm{KH}}$ the instability can grow ($\eta < \eta_{\mathrm{Crit}}$), it might eventually lead to a quasi-laminar flow. This can be seen in the top-right panel of Fig.~\ref{fig:temp_slices}, where a laminar regime has already been reached. 

In the strong cooling regime (high Da, bottom panels), the results look similar regardless of the amount of viscosity. In all cases the hot and cold medium are clearly separated and the front consists of (approximately) one cell only. Surprisingly, in contrast to what we see in the weak cooling regime, the simulations show some movement in the interface independently of the amount of viscosity. This is analyzed in Sect.  \ref{sec:strong_cooling}. 

\subsubsection{Effect of viscosity on the cooling efficiency}
The luminosity of TRMLs is predominantly due to enthalpy flux (i.e., mass conversion from hot to cold medium) and not due to dissipation of kinetic energy (see \citealp{Ji_2019, Tan_2021}). We checked that this is also true in the case with viscosity (see Appendix~\ref{app:mdot_L}), specifically also that viscous heating plays a negligible role.\footnote{We checked that viscous dissipation (RHS of Eq.~\ref{eqn:energy_eq}) contributes negligibly to $Q$ by switching off the relevant terms explicitly and found no differences in the overall cooling.\label{footnote:viscous_diss}}

To study  the difference in results with and without viscosity in a quantitative way, we need to measure the surface brightness $Q$ in each case for each $\mathrm{Da_{mix}}$. This surface brightness is calculated by measuring the total luminosity of the box $L_{\rm rad}$ and dividing it by the area of the section of the box ($Q = L_{\rm rad} / \lambda^2$).

\begin{figure*}
   \centering
   \includegraphics[width=\hsize]{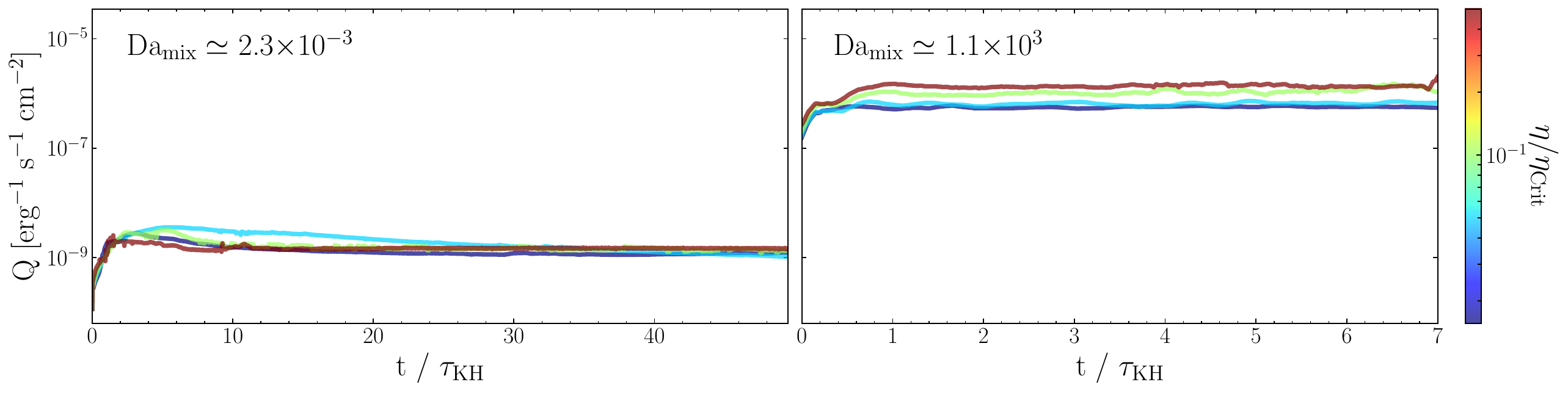}
      \caption{Evolution of the surface brightness due to cooling in the mixing layers. \textit{Left}: Evolution of $Q$ color-coded by viscosity in a very weak cooling regime with $\mathrm{Da_{mix}} \simeq 2.3\times10^{-3}$. \textit{Right}: Evolution of $Q$ for different viscosities in a very strong cooling regime with $\mathrm{Da_{mix}} \simeq 1.1\times10^{3}$.}
      \label{fig:Q_comparison}
\end{figure*}

In Fig.~\ref{fig:Q_comparison} we show the effect of viscosity for the simulations with weakest cooling (left panel) and the simulations with the strongest cooling (right panel) considered. Note that the dark blue lines correspond to the "No visc" run, considering the estimated numerical viscosity. Although visually viscosity suppresses the development of turbulence in the weakest regime (see the top panel of Fig.~\ref{fig:temp_slices}), the surface brightness obtained after several $\tau_{\mathrm{KH}}$ is very similar in all cases. 

In the case of strong cooling (right panel of Fig.~\ref{fig:Q_comparison}),\footnote{Since for $\mathrm{Da_{mix}} > 1$, $t_{\mathrm{cool}} < \tau_{\mathrm{KH}}$, we run the simulations for less Kelvin-Helmholtz times.} the differences become greater, with an increase of $Q$ with viscosity. This might be counterintuitive if we are thinking of viscosity as a force suppressing the turbulent movement and, therefore, the cooling process. This is analyzed in more detail in Sect. \ref{sec:strong_cooling}.

\begin{figure}
   \centering
   \includegraphics[width=\hsize]{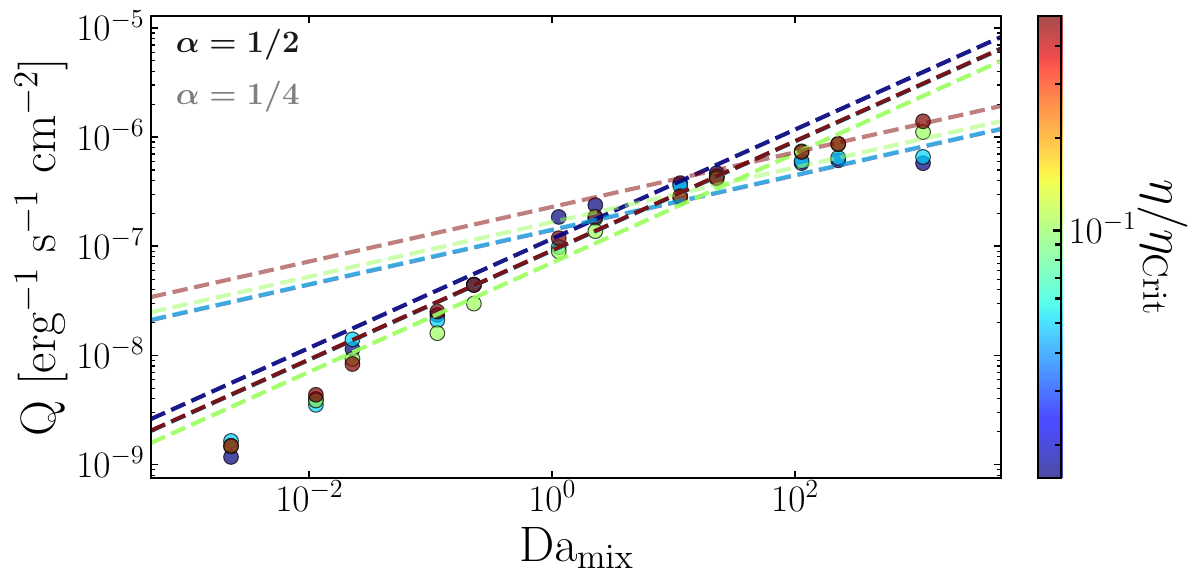}
   \caption{Surface brightness, $Q$, as a function of $\mathrm{Da_{mix}}$ color-coded by viscosity. The darker dashed line shows the best fit in each case in the weak cooling regime ($\rm Da < 1$) with a slope of $\alpha=1/2$. The lighter dashed line shows the best fit in each case in the strong cooling regime ($\rm Da > 1$) with a slope of $\alpha=1/4$.}
   \label{fig:Q_vs_Da}
\end{figure}

In Fig.~\ref{fig:Q_vs_Da} we show the relation between surface brightness and Da (Eq. \ref{eqn:Da_Q_u}). Our simulations without viscosity reproduce the expected broken power-law behavior (see Eq.~\ref{eqn:Da_Q_u}) with the change of regime from weak to strong cooling occurring at $\mathrm{Da_{mix}} \simeq 1$\footnote{The values for very weak cooling lie under the fitted line. This was also found in \citet{Das_2023}. For this reason, for the fit we take the values of $\mathrm{Da_{mix}} > 5\times10^{-2}$.}. Although in the weak cooling regime, the differences are small, in the strong cooling regime viscosity leads to larger values of $Q$. This implies that the transition from weak to strong cooling regime is shifted toward larger values of $\mathrm{Da_{mix}}$ when increasing the amount of viscosity. 

In summary, both Figs.~\ref{fig:Q_comparison} and \ref{fig:Q_vs_Da} reveal two counterintuitive behaviors: while we expect viscosity to suppress the mixing process and, thus, the amount of intermediate temperature gas available for cooling, we find \textit{(i)} for the weak cooling regime a mass transfer rate between the phases independent of viscosity, and \textit{(ii)} for the strong cooling regime even an increased $\dot m$ with larger values of viscosity.
This implies that either the naive expectation of how viscosity affects turbulence is wrong, or the $u'$-$Q$ relation is changed due to viscosity. We went on to explore this by analyzing the turbulence inside the mixing layer in the next sections.

\subsubsection{Effect of viscosity on the turbulence}

From visual inspection of Fig.~\ref{fig:temp_slices}, viscosity seems to suppress turbulence in the weak cooling regime, but not in the strong cooling regime. To test this in a quantitative way, we computed the turbulent velocity in our domain following \citet{Tan_2021}; namely, to avoid the effect of bulk motions, we measured the velocity dispersion in the $\hat{z}$ direction, by taking the maximum of $v_z$ profile for each snapshot and then averaging over several $\tau_{\rm KH}$\footnote{In this process, we also check the velocity dispersion in the $\hat{x}$ and $\hat{y}$ direction after subtracting the bulk motions, making sure that turbulence is isotropic.}.

\begin{figure}
   \centering
   \includegraphics[width=\hsize]{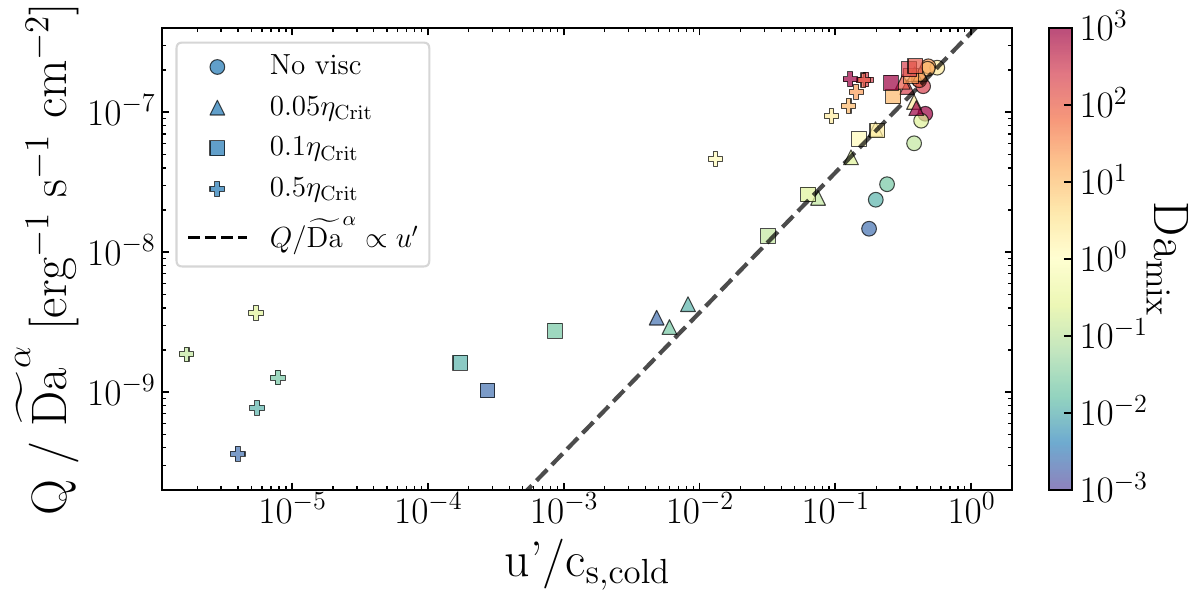}
   \caption{Relation between $Q$, normalized by the numerical measured Damköhler number ($\widetilde{Da}$) in each regime, and the turbulent velocity for different viscosities and color-coded by $\mathrm{Da_{mix}}$. The dashed line shows the relation $Q/\widetilde{Da}^{\alpha} \propto u'$ expected.}
   \label{fig:u_turb_vs_Q}
\end{figure}

Figure~\ref{fig:u_turb_vs_Q} shows relation \ref{eqn:Da_Q_u}, between $Q$ and the turbulent velocity normalized to the cold gas sound speed, color-coded by $\mathrm{Da_{mix}}$. To get rid of the $\mathrm{Da_{mix}}$ dependence, we normalize the value of $Q$ to the value of $\widetilde{Da}^{\alpha}$, as done in \citet{Das_2023}. $\widetilde{Da}$ is calculated from Eq. \ref{eqn:Da_theory}, using the turbulent velocity measured from the simulations and $\alpha$ corresponds to the index from Eq. \ref{eqn:Da_Q_u}. By doing this normalization, $Q / \widetilde{Da}^{\alpha} \propto u'$. This relation is satisfied in the strong cooling regime ($\mathrm{Da_{mix}} > 1$), regardless of the amount of viscosity. However, in the weak cooling regime ($\mathrm{Da_{mix}} < 1$), the relation is satisfied for the "No visc" (circles) and $0.05\eta_{\mathrm{Crit}}$ (triangles) cases. The turbulent velocity of the runs with $0.1\eta_{\mathrm{Crit}}$ (squares) are one order of magnitude lower than expected and the runs with $0.5\eta_{\mathrm{Crit}}$ (crosses), more than two orders of magnitude lower.

This result indicates that, as previously noted by visual inspection in Fig.~\ref{fig:temp_slices}, viscosity suppresses turbulence, but it does not reduce surface brightness. However, the suppression occurs in the weak cooling regime, whereas in the strong cooling regime, the relation between $Q$ and $u'$ is not modified. This is surprising, due to the weak dependence of $u'$ on $\mathrm{Da}$ ($u' \propto \rm Da^{1/10}$; see \citealp{Tan_2021}).

\begin{figure}
   \centering
   \includegraphics[width=\hsize]{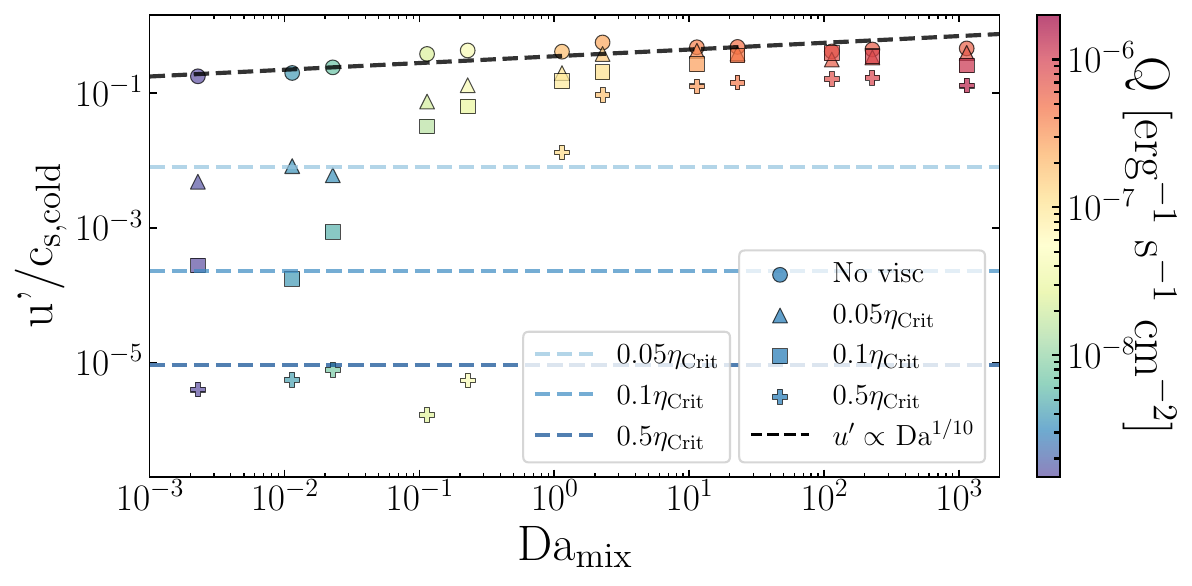}
   \caption{Dependence of the turbulent velocity on $\mathrm{Da_{mix}}$ for different amounts of viscosity and color-coded by the surface brightness. The dashed black line indicates the dependency of $u' \propto \mathrm{Da_{mix}}^{1/10}$. The dashed blue lines show the turbulent velocity values of the adiabatic system with different viscosities.}
   \label{fig:u_turb_vs_Da}
\end{figure}

We also tested this weak dependence of the cooling on turbulence explicitly and give the results in Fig.~\ref{fig:u_turb_vs_Da}.
The simulations without viscosity fall on the $u' \propto \mathrm{Da}^{1/10}$ relation found in \citet{Tan_2021}. In the weak cooling regime (${\rm Da_{mix}}<1$) the relation is not satisfied in the viscous runs.
The turbulent velocities drop until they reach the $u'$ in adiabatic simulations, saturating at this value (dashed blue lines). The adiabatic system sets the floor for turbulent velocities depending on viscosity.

In this weak cooling regime, the simulations show the expected behavior: the more viscous the system is, the less turbulent it becomes. However, for larger values of $\mathrm{Da_{mix}}>1$, all the runs tend to follow the $u' \propto \mathrm{Da}^{1/10}$; namely, the runs mostly display a level of turbulence \textit{(a)} that is much higher than expected and \textit{(b)} fairly independent of viscosity. 

Furthermore, Fig.~\ref{fig:u_turb_vs_Da} also shows that the characteristic Damköhler number at which the simulations follow the expected $u'\propto {\rm Da}^{1/10}$ relation increases with increasing viscosity (i.e., the runs with lower viscosity start following the relation at a lower $\mathrm{Da_{mix}}$), while the more viscous runs occur at a higher $\mathrm{Da_{mix}}$. In the strong cooling regime, all the simulations follow the $u' \propto \mathrm{Da}^{1/10}$ relation, suggesting that the effect of viscosity is reduced due to cooling.

Viscosity acts as a friction, reducing the velocity of the gas and, therefore, it is expected to produce lower values of turbulent velocity. However, two questions arise, which we explore in the following subsections.  
\begin{enumerate}
    \item Viscosity suppresses turbulence in the weak cooling regime ($\rm Da< 1$) -- as expected. However, we need to investigate why the amount of cooling is only weakly affected (see Fig.~\ref{fig:Q_vs_Da}), thereby effectively breaking the universal $u'-Q$ relation (see Fig.~\ref{fig:u_turb_vs_Q}).
    \item On the other hand, viscosity has a weak effect on both the level of turbulence as well as the amount of cooling in the strong cooling regime (see Figs.~\ref{fig:u_turb_vs_Q} and \ref{fig:u_turb_vs_Da}). This is particularly puzzling, since turbulence only depends very weakly on cooling (see Fig.~\ref{fig:u_turb_vs_Da}). Thus, even with very strong cooling turbulence should not directly "overpower" the effect of viscosity.
\end{enumerate}

\subsubsection{Weak cooling regime}
To address the first question, namely, why for $\rm Da < 1$ the cooling rate is not affected by viscosity (but $u'$ is), we need to understand the behavior of a viscous gas in the weak cooling regime. Due to the long cooling time, the gas behaves similarly to an adiabatic gas, which has been previously studied in detail \citep[e.g.,][]{Roediger_2013, Marin-Gilabert_2022}. Viscosity smooths out the shear velocity gradient and suppresses the growth of instabilities, leading to a reduced turbulent velocity (see the top panel of Fig.~\ref{fig:vx_profile}). However, if turbulence is not the driver for mixing, we must explore why $Q$ is on the same order of magnitude regardless of the amount of viscosity. 

When viscosity is strong enough, the instabilities are suppressed, leading to a laminar flow, rather than a turbulent flow. The theoretical relation in Fig.~\ref{fig:u_turb_vs_Q} assumes turbulent mixing \citep[e.g.,][]{Damkohler_1940, Shchelkin_43}, therefore it is not surprising that for high viscosities the simulations do not follow the expected relation. Viscous diffusion dominates over convection and cooling, increasing the thickness of the interface, producing that the theoretical analysis fails \citep[e.g.,][]{Klimov_1963, Libby_1982}. Although the viscous runs present a laminar flow, the amount of gas at intermediate temperatures where the cooling curve peaks is similar among the viscous cases (see the top row of Fig.~\ref{fig:temp_hist}). The nonviscous case leads to a higher amount of intermediate temperature gas due to turbulent mixing. However, the long cooling time in the weak cooling regime produces a bottleneck effect, which explains the constant $Q$ obtained, despite the different diffusion mechanisms. Note that our criterion for critical viscosity is defined within $1\tau_{\rm KH}$, but our simulations in the weak cooling regime are run until $50\tau_{\rm KH}$. This means that, although within $1\tau_{\rm KH}$ the instability might grow, the system might become laminar after several $\tau_{\rm KH}$.

\subsubsection{Strong cooling regime} \label{sec:strong_cooling}

To answer the second question, namely, why neither $u'$, nor the cooling are strongly affected by viscosity for $\rm Da>1$, we need to study the strong cooling regime in detail. Viscosity is expected to reduce turbulence, although in the strong cooling regime, the $Q - u'$ relation follows the analytical expectations from turbulent theory. Thus, if viscosity is indeed expected to suppress turbulence, we must consider why $u'$ is similar in all cases regardless of the amount of viscosity. 

In the strong cooling regime, cooling acts much faster than turbulent mixing and ends up quickly cooling  all the intermediate gas and producing  a front where cooling occurs that is $\sim 1$ cell thick. As a consequence, cooling tends to keep the temperature profile sharp and, if the system is in pressure equilibrium, also the momentum and shear velocity profiles. This is the opposite effect to viscosity, which tends to smooth out the shear velocity gradient. This produces that, in the strong cooling regime, cooling dominates over viscosity, keeping the shear velocity profile sharp and, as a consequence, being able to induce turbulence in a similar way as the nonviscous case and reducing the effect of viscosity (see Fig.~\ref{fig:vx_profile}). 

\begin{figure}
   \centering
   \includegraphics[width=\hsize]{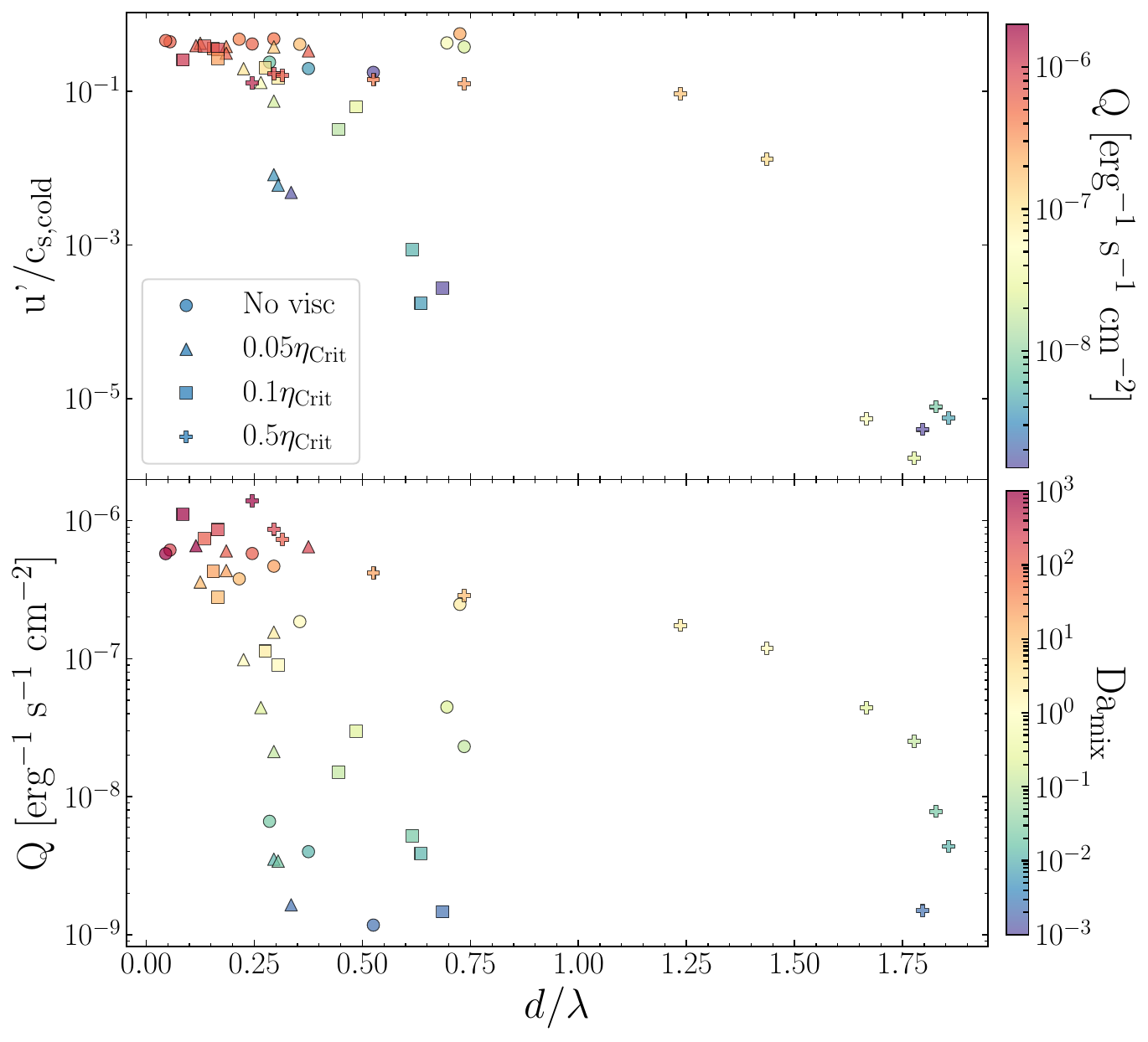}
   \caption{\textit{Top}: Turbulent velocity as a function of the velocity gradient width, color-coded by the surface brightness. \textit{Bottom}: Surface brightness as a function of the velocity gradient width, color-coded by ${\rm Da_{mix}}$.}
   \label{fig:d_shear}
\end{figure}

Keeping the shear profile sharp leads to continuous KHIs forming, and hence to turbulence, as we can see in the top panel of Fig.~\ref{fig:d_shear}, where we show the dependence of turbulence on the scale $d$ at which the shear velocity profile is smoothed after $t = 1.25\tau_{\rm KH}$ (see Appendix \ref{app:d_shear} for details). For "No visc" and $0.05\eta_{\rm Crit}$, $d$ is always small, producing larger values of $u'$ (i.e., turbulence). For the cases of $0.1\eta_{\rm Crit}$ and $0.5\eta_{\rm Crit}$, the weak cooling regime results show the effect of viscosity in smoothing out the gradient (large values of $d$), leading to a low $u'$. However, in the strong cooling regime the shear velocity gradient becomes sharp (low values of $d$), inducing turbulence. 
In summary, while cooling does not (strongly) affect the turbulence directly, it sharpens the density gradient required to seed KHI, and hence combats the smoothing effect of viscosity.

In other words, a sharp velocity gradient induces turbulence and, therefore, leads to larger surface brightness. This can be seen in the bottom panel of Fig.~\ref{fig:d_shear}, where larger values of $\rm Da_{mix}$ lead to sharper gradients (lower $d$) and, therefore, larger $Q$.\\

To test our assumption that the system follows the expected $u' \propto \mathrm{Da}^{1/10}$ when cooling dominates over viscosity, we estimate analytically at which value of ${\rm Da_{mix}}$ this should happen.
Cooling dominates over viscosity when $t_{\rm cool} < \tau_{\nu}$, where $t_{\rm cool}$ is measured at $T_{\mathrm{mix}} = 2\times10^5$~K and $\tau_{\nu}$ corresponds to the time that a given viscosity takes to smooth out the velocity gradient a distance $d = \lambda/10$ (see Sect. \ref{sec:theoretical_estimate}).

In our case, we express the different amounts of viscosity as a function of the critical viscosity as $\eta = \zeta \, \eta_{\rm Crit}$, where $\zeta$ is the fraction of the critical viscosity used. Using the equation for critical viscosity (Eq. \ref{eqn:nu_crit}), KH time (Eq. \ref{eqn:tau_KH}), and $\nu = \eta / \rho$, we can express $\eta$ as
\begin{equation}
    \eta = \zeta \, \eta_{\rm Crit} = \zeta \frac{\lambda^2 \rho}{100 \tau_{\rm KH}} \, .
\end{equation}
We can also express $t_{\rm cool}$ as
\begin{equation}
    t_{\rm cool} = \frac{\tau_{\rm KH}}{{\rm Da}} \, ,
\end{equation}
where $t_{\rm turb} = \lambda/u' \sim \tau_{\rm KH}$ and the viscous timescale as
\begin{equation}
    \tau_{\nu} = \frac{d^2}{\nu} = \frac{d^2 \rho}{\eta} = \frac{\lambda^2 \rho}{100 \, \eta} = \frac{\tau_{\rm KH}}\zeta \, .
\end{equation}

\noindent Therefore, cooling will dominate over viscosity when

\begin{equation}
    {\rm Da} > \zeta \, .
\end{equation}

\begin{figure}
   \centering
   \includegraphics[width=\hsize]{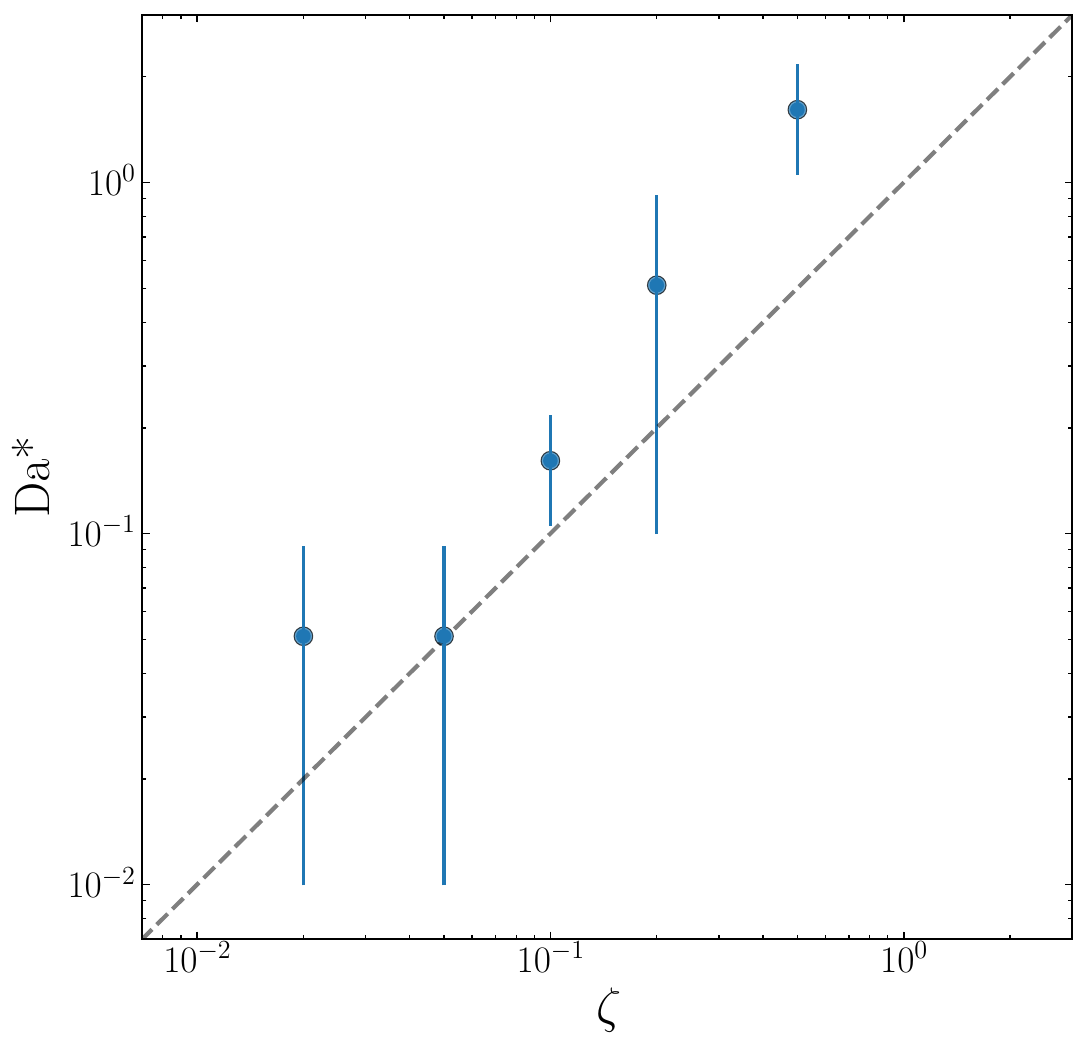}
   \caption{Measured critical Da ($\rm Da^*$) vs. the fraction of critical viscosity ($\zeta$) for different sets of simulations. We included two more sets for completeness that are not shown in the previous plots. The dashed line indicates the expected one-to-one relation.}
   \label{fig:K_vs_Da_star}
\end{figure}

We tested this estimate with our simulations by checking the critical value of ${\rm Da_{mix}}$ (which we denote as ${\rm Da^*}$), when the simulations start to follow the $u' \propto \mathrm{Da}^{1/10}$ relation for the different amounts of viscosity (see Fig.~\ref{fig:u_turb_vs_Da} in which can be read off of ${\rm Da^*}$ directly). We included two more sets of simulations for completeness and we plot the computed values of ${\rm Da^*}$ as a function of $\zeta$ in Fig.~\ref{fig:K_vs_Da_star}. 

Although the theoretical estimate was based on broad assumptions, the computed values of ${\rm Da^*}$ from our simulations fit the theoretical estimate. This agrees with our previous assumption that cooling effectively reduces the effect of viscosity by keeping the temperature gradient sharp.

To understand why cooling is increased (by a factor of $\sim 2$) for the more viscous runs at $\rm Da > Da^*$, we start by recalling that while cooling reduces the effect of viscosity, the viscous runs still lead to a broader front (see the bottom row of Fig.~\ref{fig:temp_slices}). As a consequence, the amount of gas with an intermediate temperature is larger with viscosity (see the bottom row of Fig.~\ref{fig:temp_hist}). This produces that the overall amount of cooling is larger with viscosity than without viscosity, leading to a slightly higher $Q$ in the strong cooling regime in the cases with higher viscosities.\footref{footnote:viscous_diss} This can also be seen in the bottom panel of Fig.~\ref{fig:d_shear} where, despite $d$ becoming small in the strong cooling regime regardless of the amount of viscosity, it is still larger in the most viscous runs, leading to a larger $Q$ for a given ${\rm Da_{mix}}$.

\begin{figure}
   \centering
   \includegraphics[width=\hsize]{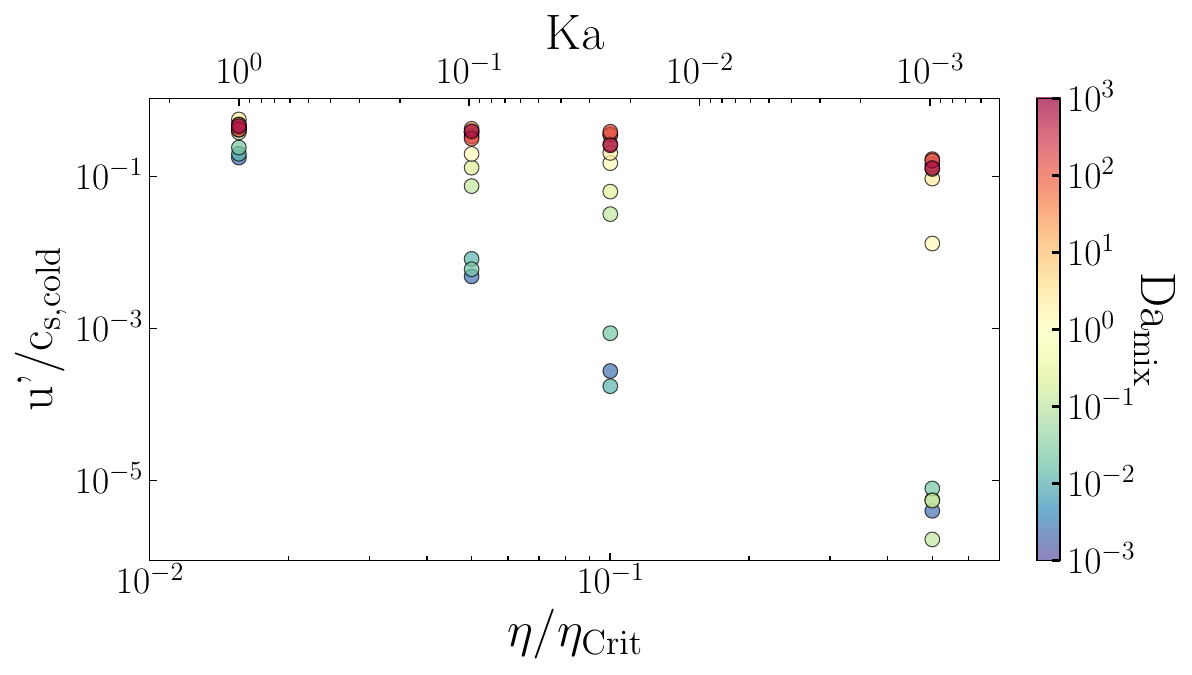}
   \caption{Turbulent velocity as a function of the amount of viscosity of the medium (Karlovitz number), color-coded by $\mathrm{Da_{mix}}$.}
   \label{fig:u_turb_vs_visc}
\end{figure}

To study how $u'$ depends on viscosity in more detail, Fig.~\ref{fig:u_turb_vs_visc} shows how turbulent velocities change with viscosity. In the nonviscous case, the data points cluster together, corresponding to the weak dependence of $u'$ on $\rm Da$. For higher viscosities, the two regimes defined by $\rm Da^*(\eta)$ can be identified in each case: the data points group separately in the viscous-dominated regime (${\rm Da_{mix}} < \zeta$) and in the cooling-dominated regime (${\rm Da_{mix}} > \zeta$).

\begin{figure}
    \centering
   \includegraphics[width=0.95\hsize]{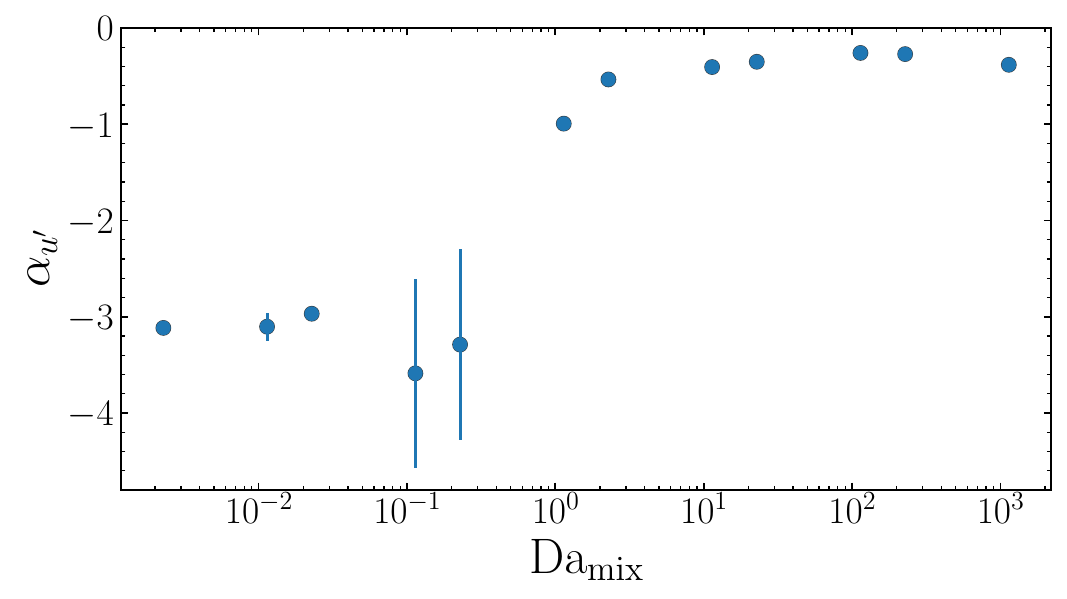}
   \caption{Exponent of the dependence of the turbulent velocity on the amount of viscosity (Karlovitz number) calculated from Fig. \ref{fig:u_turb_vs_visc} as a function of $\mathrm{Da_{mix}}$.}
   \label{fig:Index_uturb_vs_visc}
\end{figure}

Assuming that $u'$ depends on viscosity by a power law $u' \propto \eta^{\alpha}$, we can fit the slopes for individual $\rm Da_{mix}$ in Fig.~\ref{fig:u_turb_vs_visc} and check the two different regimes. Figure~\ref{fig:Index_uturb_vs_visc} shows the different values of $\alpha$ depending on $\rm Da_{mix}$, reflecting the change of regime from viscous-dominated to cooling-dominated at $\rm Da_{mix} \sim 0.5$. In the viscous-dominated regime (low $\rm Da_{mix}$) the turbulent velocity decays with viscosity as $u' \propto \eta^{-3}$, while in the cooling-dominated regime (high $\rm Da_{mix}$) the exponent is close to zero, indicating that the system behaves similarly regardless of the amount of viscosity\footnote{Since the change of regime depends on the value of viscosity, the ``transition zone'' produces significant errors in the fit, where low-viscosity runs already behave  as if they were nonviscous, but high-viscosity runs are still viscous-dominated.}.

\section{Discussion} \label{sec:discussion}

\subsection{Temperature-dependent viscosity}
\label{sec:hot_vs_cold_viscosity}

Considering a more realistic Spitzer viscosity dependent on the temperature of the gas, each fluid will have a viscosity $\eta$. In the case of the cold fluid: $\eta_{\rm Cold} \propto T_{\rm Cold}^{5/2}$; while in the hot medium: $\eta_{\rm Hot} \propto T_{\rm Hot}^{5/2}$. Therefore, by comparing both viscosities, we have
\begin{equation}
    \eta_{\rm Hot} = \eta_{\rm Cold} \left(\frac{T_{\rm Hot}}{T_{\rm Cold}} \right)^{5/2} = \eta_{\rm Cold} \, \chi^{5/2} \, .
\end{equation}
In the case of an overdensity of $\chi = 100$: $\eta_{\rm Hot} = 10^5 \, \eta_{\rm Cold}$, rendering the viscosity of the cold medium negligible.

Additionally, the diffusivity of a system depends on the diffusion coefficient, which in the viscous case is set by the kinematic viscosity ($\nu$). For a system with a constant dynamic viscosity ($\eta$, as in our setup) but two different densities, each fluid will have a different diffusivity (see Eq. \ref{eqn:kinematic_coeff}). Since the kinematic viscosity depends on the inverse of the density, for a given dynamic viscosity, the cold medium will have a lower diffusivity compared to the hot medium. As a consequence, for systems with a high density contrast, the diffusivity of the system is set by the hot gas. 

\begin{figure}
    \centering
   \includegraphics[width=\hsize]{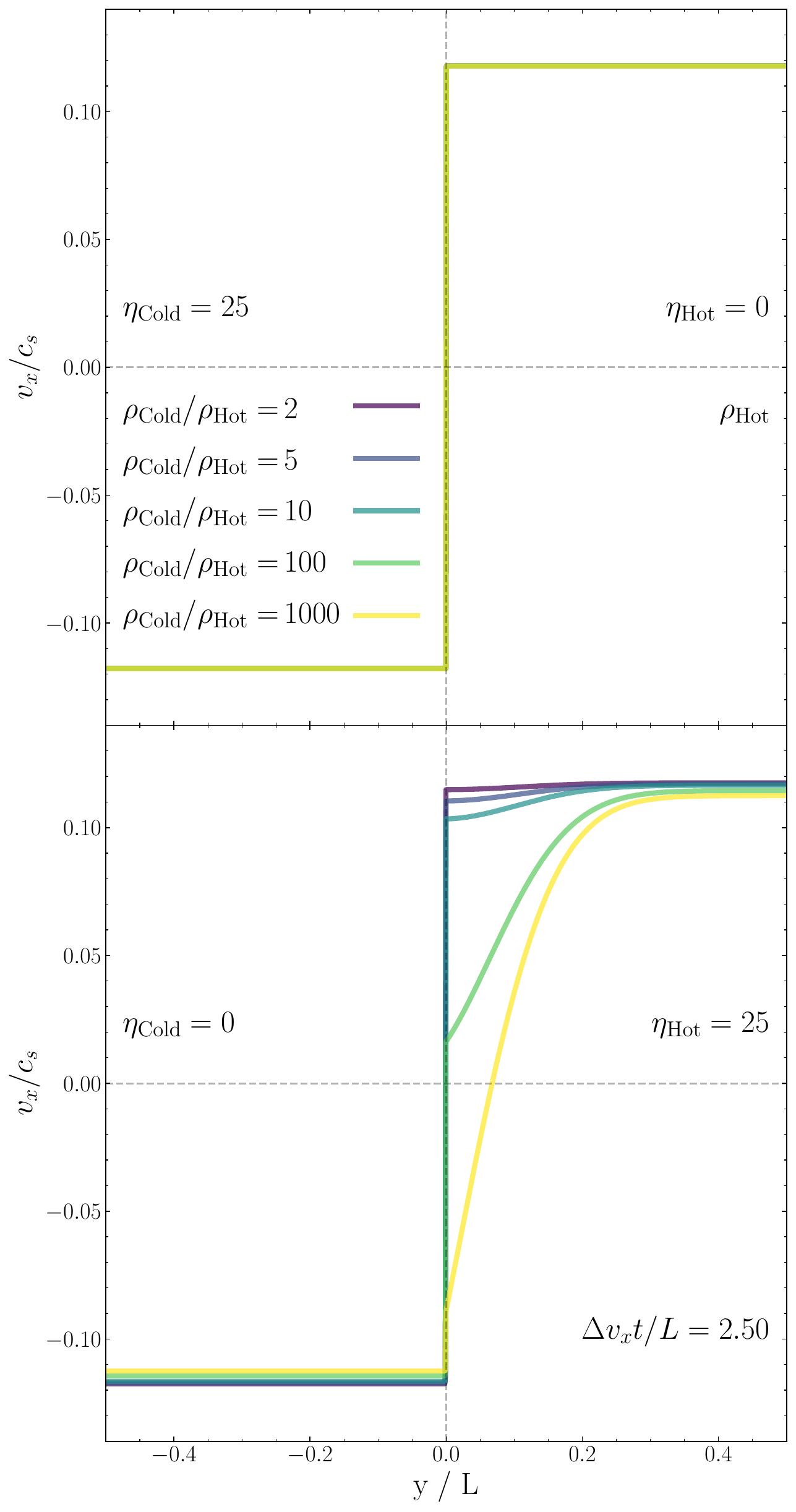}
   \caption{Velocity profile of 1D simulations. {\it Top}: Only the cold gas ($y < 0$) is viscous. {\it Bottom}: Only the hot gas ($y > 0$) is viscous.}
   \label{fig:hot_vs_cold_visc}
\end{figure}

Taking into account that: with a more realistic temperature-dependent Spitzer viscosity, the dynamic viscosity of the cold medium is negligible and that the viscous diffusion is set by the hot gas, a viscous cold medium should have a small impact compared to a viscous hot medium. We tested this using 1D simulations, with the following setup:
\begin{equation}
(\rho,\, T,\, v,\, \eta) = 
\begin{cases}
    (\rho_{\rm Cold}, \, T_{\rm Cold}, \, -v_x, \, \eta_{\rm Cold}) & \text{for } y < 0, \\
    (\rho_{\rm Hot}, \, T_{\rm Hot}, \, v_x, \, \eta_{\rm Hot}) & \text{for } y > 0 \, ,
\end{cases}
\end{equation} 
where $\Delta v_x/c_s = 0.24$, we keep $T_{\rm Hot}$ and $\rho_{\rm Hot}$ constant and change density (and temperature) of the cold medium as $T_{\rm Hot} / T_{\rm Cold} = \rho_{\rm Cold} / \rho_{\rm Hot} = 2, 5, 10, 100, 1000$. To analyze the diffusivity depending on the medium, we set a constant viscosity of $\eta = 25$ (in internal units) first in the cold medium and then in the hot medium. Figure~\ref{fig:hot_vs_cold_visc} shows the $v_x$ profile after $\Delta v_x t / L = 2.50$ for the two cases: the viscous cold medium on the top panel and the viscous hot medium on the bottom panel.

When the cold medium is viscous (top panel in Fig.~\ref{fig:hot_vs_cold_visc}), diffusivity is decreased as the cold medium becomes denser.  
Thus, if the initial (i.e., lowest overdensity) value of viscosity is too low to have a discernible effect, it becomes increasingly negligible at higher overdensity. 
When the hot medium is viscous (bottom panel in Fig.~\ref{fig:hot_vs_cold_visc}), diffusivity remains constant, since the density of the hot medium is kept constant. However, the velocity profile becomes smoother as the cold medium becomes denser, despite the cold medium not being viscous. This happens because, although the cold medium is not viscous, it carries a lot more momentum than the hot (and viscous) medium. Therefore, when the cold medium transfers momentum to the hot medium, viscosity changes the momentum of the hot medium very effectively, smoothing the velocity gradient. The denser the cold medium is, the more momentum is transferred to the hot medium, leading to a smoother velocity profile. These results corroborate the assertion reported in Sect. \ref{sec:crit_visc_overdens}, namely, that the hot phase dominates viscosity and the cold phase dominates momentum. 
We note that this leads to the interesting effect that higher overdensities imply a larger degree of smoothing of the velocity profile, however, as $\eta_{\rm crit}$ also rises (see Eq.~\ref{eqn:critical_eta} and Fig.~\ref{fig:crit_visc_chi}), a higher viscosity is required to suppress the KHI.

Considering a constant $\eta$ for the whole system leads to a diminishing momentum diffusion in the cold medium, similarly to a realistic temperature-dependent Spitzer viscosity. Therefore, a constant $\eta$ would lead to more realistic results than considering a constant $\nu$, where both media would be equally diffusive.

\subsection{Super-critical viscosity}
In this paper, we  focus on the sub-critical viscosity regime, where all the cases studied have a viscosity smaller than the critical viscosity. The instability is able to grow within one Kelvin-Helmholtz  time, being enhanced by strong cooling and allowing to keep the turbulence (see Sect.  \ref{sec:strong_cooling}). 

With a super-critical viscosity ($\eta > \eta_{\rm Crit}$) the instability is fully suppressed and there is no initial growth. This means that cooling cannot enhance the growth of the instability and dominate over viscosity. To test this, we ran a simulation with $\eta = 2\eta_{\rm Crit}$ at $\rm Da_{mix} \simeq1.1\times10^{2}$, which is an order of magnitude larger than the estimated Damköhler number required to overpower viscosity $\rm Da^*$. Since the viscous length is larger than the wavelength for super-critical viscosity, we increase the size of the box to $L_x = L_z = 3\lambda$. 

\begin{figure}
    \centering
   \includegraphics[width=\hsize]{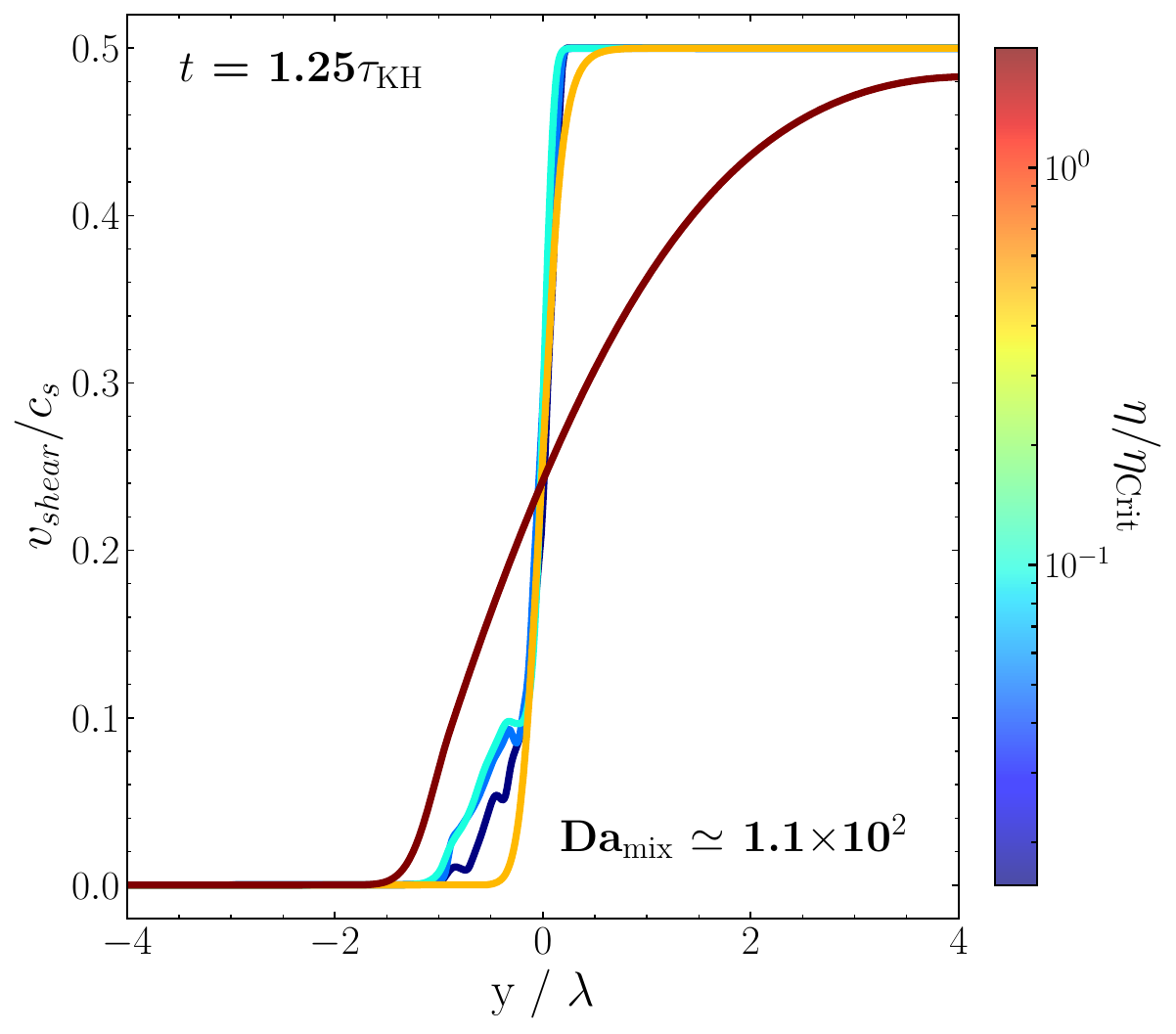}
   \caption{$v_{\rm shear}$ profiles at $\rm Da_{mix} \simeq1.1\times10^{2}$ and $t = 1.25\tau_{\rm KH}$ including the run with $2\eta_{\rm Crit}$. The distance, $d$, at which the shear profile has been smoothed in the $2\eta_{\rm Crit}$ case is large compared to the other cases.}
   \label{fig:vx_profile_2eta}
\end{figure}

In Fig.~\ref{fig:vx_profile_2eta} we show the shear velocity profile after $t = 1.25\tau_{\rm KH}$ for different levels of viscosity. For $\eta < \eta_{\rm Crit}$ the effect of cooling keeps the velocity profile sharp, while for $\eta = 2\eta_{\rm Crit}$ viscosity is able to smooth out the gradient, although $\rm Da_{mix} \gg Da^*$.

As a consequence, the turbulent velocity is suppressed and viscosity still dominates over cooling. The turbulent velocity of the run with $2\eta_{\rm Crit}$ is $u'/\rm{c_{s,cold}} = 9.4\times10^{-7}$, behaving as viscous-dominated (see Fig.~\ref{fig:u_turb_vs_Da} for reference).

When $\eta < \eta_{\rm Crit}$, there is an initial growth of the KHI that is later enhanced by cooling, leading to a cooling-dominated regime where the system behaves as nonviscous. However, if $\eta > \eta_{\rm Crit}$ there is not initial growth that can be enhanced by cooling, therefore the system will always be viscous-dominated.

In the temperature domain we consider in this paper, where the cooling function peaks ($10^4~{\rm K} < T < 10^6$~K), the Spitzer value for the $2 \eta_{\rm Crit}$ is $2 \eta_{\rm Crit} \simeq 7.10 \eta_{\rm Sp}$ (see Sect. \ref{sec:TRMLs} and Table~\ref{tab:TRMLs_runs}), which is not realistic in plasma physics \citep[][]{Kunz_2022}. Therefore, an astrophysical scenario with such a high level of viscosity is very unlikely.

\subsection{Applications}

Viscosity becomes relevant in very hot and ionized media, with temperatures typically above $10^6$~K. On the other hand, cooling efficiency peaks between $10^4-10^6$~K. This means that both cooling and viscosity will act together in astrophysical scenarios where very hot and cold media co-exist and mix, which is the case for the ICM, ISM, and CGM -- but also in other multiphase systems such as galactic winds or coronal rain. This interaction might happen in different astrophysical processes, where the Reynolds number is comparable to the ones used in this paper (see Table \ref{tab:TRMLs_runs}).
Similarly, mixing and radiative cooling are physical processes that occur frequently in these multiphase systems. In fact, viscosity, mixing and cooling are commonly thought of as being crucial in determining their long-term evolution. Examples of such multiphase systems include: 
\begin{itemize}
\item Galaxies moving through the ICM experience a ram pressure that strips the cold ISM \citep[e.g.,][]{Gisler_1976, Nulsen_1982}. The growth of instabilities in the tails mixes the hot ICM with the cold ISM \citep[e.g.,][]{Iapichino_2008, Gosh_2024}. As a consequence, the hot ICM cools down, providing new gas and enhancing star formation \citep{Muller_2021}. Viscosity not only affects the morphology and mixing of the tail with the medium \citep[e.g.,][]{Kraft_2017, Marin-Gilabert_2024}, but also the cooling efficiency and, therefore, it might change the star formation rate.
\item The existence of multiphase gas within galactic winds is commonly often attributed to the effects of continuous cooling (see the reviews by \citealt{Veilleux_2020} and \citealt{Thompson_2024}). 
Similarly, cold clouds infalling through the CGM is another example where the hot gas of the CGM mixes with the cold gas of the cloud being disrupted \citep{Gronke_2018, Gronke_2022}. When the CGM is hot enough ($T > 10^6$~K), transport processes become relevant and shape the morphology of the tails of the cloud \citep{Bruggen_2023}. Viscosity affects the cloud dynamics, as well as the evolution and evaporation of the tail \citep{Jennings_2021}. Therefore, it also affects the amount of cold gas and the mixing efficiency of the cloud and the CGM. The mixing between the cold stripped gas and the medium can be observed and studied using different ion lines \citep[e.g.,][]{Kwak_2010}. 
\item Cold streams in hot galactic halos are likely crucial to our understanding of galaxy fueling and growth \citep{Dekel_2009}. As other classical example of multiphase gas interactions, mixing, cooling, and the effect of viscosity are discussed here in the context of cold streams \citep{Bruggen_2005, Mandelker_2020}.
\end{itemize}

In each of these systems, turbulence, mixing, and cooling are crucial to understanding the dynamics and ultimate fate of the gas. As a result, key findings from studies of TRMLs \citep[e.g.,][]{Ji_2019, Fielding_2020, Tan_2021} are commonly invoked to interpret a wide range of observational diagnostics \citep[e.g.,][]{Qu_2022, Lin_2025}. However, most previous works have neglected the role of viscosity, meaning that such interpretations could be significantly biased; in particular, if viscosity were to strongly suppress turbulent mixing and, consequently, cooling, as might be naively expected.

In this study, however, we show that (contrary to expectations) viscosity has only a modest impact on the dynamics of turbulent mixing layers. In particular, for systems in the fast cooling regime (Da~$>1$), which includes most of the astrophysical environments described above, not only does the $u’$–$Q$ relation remain unchanged, but also the amount of turbulence expected from KHI. This allows us to continue using established theoretical predictions to relate the line widths to expected emissivities and other properties.\\

Another important application of our work relates to the growing number of multiphase subgrid models, which aim to address the long-standing resolution problem in large-scale (e.g., cosmological) simulations. Such simulations are unable to fully resolve multiphase structures, such as those discussed above, and therefore rely on sub-grid prescriptions to approximate their behavior \citep[][]{Huang_2020, Smith_2023, Butsky_2024, Das_2024}. These models commonly adopt a numerical multi-fluid framework \citep[e.g.,][]{Weinberger_2022}, with source and sink terms for each fluid component informed by small-scale simulations of  turbulent mixing layers \citep[e.g.,][]{Tan_2021}.

However, to date, none of these models have accounted for the combined effects of radiative cooling and viscosity. Instead, they implicitly assume that the omitted effect (e.g., viscosity) has a negligible influence on the relevant dynamics. In this work, we provide a robust physical foundation for current and future implementations of multiphase sub-grid models by explicitly testing this assumption. In particular, we show how mass transfer rates between phases are (largely) unaffected by viscosity, thus justifying the use of existing cooling-based prescriptions in multi-fluid models.\\

Taken together, our results show that viscosity, while potentially important in shaping large-scale morphology, does not significantly alter the fundamental scaling relations governing turbulent mixing and radiative cooling. Although we might naively expect viscosity to suppress mixing and thereby invalidate commonly used theoretical predictions, we find (perhaps surprisingly) that this is not the case. This lends confidence to both the interpretation of observational diagnostics and the construction of multiphase subgrid models, which often rely on idealized, inviscid small-scale simulations. By clarifying the limited role of viscosity, our work helps bridge the gap between detailed local physics and the broader astrophysical systems in which these processes operate.

\subsection{Caveats and future work}

This paper focuses on a very idealized setup for simplicity, to isolate the effect of viscosity and cooling. However, future projects can expand this work further, for instance, by focusing on:

\begin{itemize}
    \item Temperature-dependent viscosity: In this study, we looked at the physical effects at play and, thus, we used a constant viscosity in our simulations for simplicity. However, Spitzer viscosity is expected to depend on the temperature of the medium, leading to a more complex system. Although viscosity is expected to be low in the cold medium, the momentum transfer from the cold and denser medium to the viscous hot medium can lead to non-negligible effects (see Sect.  \ref{sec:hot_vs_cold_viscosity}). Additionally, the mass inflow due to cooling might also be affected, since gas with different temperatures will have different viscosities. Temperature-dependent viscosity could also potentially affect the steady-state amount of gas at different temperatures (i.e., the temperature probability density function), similarly to how the temperature dependence of thermal conduction strongly affects the temperature probability density function (PDF) \citep{tan21-lines}. The temperature PDF is crucial to predicting absorption and emission line ratios. Interestingly, the absolute value of viscosity appears to affect the temperature PDF (Fig.~\ref{fig:temp_hist}), whereas the absolute value of conduction does not appear to affect the temperature PDF, which only depends on how conduction scales with temperature \citep{tan21-lines}. Further investigations of such effects are beyond the scope of this paper.
    \\
    \item Anisotropic viscosity: In magnetized media such as the CGM, where the particles' gyro-radius is small compared to the mean free path, the effect of viscosity is expected to be anisotropic. This leads to a lower viscosity perpendicular to the magnetic field lines, decreasing the total effect of viscosity. A more realistic setup should include magnetic fields, which can suppress the growth of instabilities \citep{Das_2023}, and an anisotropic viscosity, which will be dependent on the magnetic field direction \citep{Zuhone_2015, Berlok_2019, Marin-Gilabert_2025b}. The combination of both might modify the morphology of the interface, depending on the direction of the magnetic field, potentially changing the cooling efficiency of the system. 
    \\
    \item Larger-scale simulations: Here, we focused on an individual mixing layer. However, more complex multiphase setups, such as the "cloud crushing" setup, exhibit turbulent mixing in several locations. Viscosity affects the growth of instabilities, which is essential for determining the survival of a cold cloud \citep[e.g.,][]{Klein_1994} subject to a uniform wind. However, it also affects cooling, which is essential in the survival and the mass growth of the cloud in a radiative turbulent medium \citep[e.g.,][]{Gronke_2018, Kanjilal_2020}. The change in morphology and cooling efficiency might lead to different criteria to determine the survival of a cloud or the mass growth. 
    We note that \citet{Li_2019} conducted such "cloud crushing" experiments including viscosity. They found a prolonged lifetime of the cloud; however, they did not study the (potential change of the) mass transfer rate between the phases in detail.\\

    \item Effect of thermal conduction: While thermal conduction is generally subdominant over turbulent diffusion and, thus, it is not thought to affect the overall level of cooling found in TRMLs \citep{Tan_2021}, the inclusion of conduction can shift the Karlovitz number (see Sect. \ref{sec:introduction}) to $\rm Ka>1$. This might potentially affect the microphysics of the cooling front. Hence, we want to include both viscosity and conduction in the future to put the mass transfer scalings on solid ground.
    \end{itemize}

\section{Conclusions} \label{sec:conclusions}

The role of viscosity in TRMLs is key for understanding the suppression of hydrodynamical instabilities and its effect on multiphase gas evolution in astrophysical environments. In this work, we investigated the impact of viscosity on the KHI and its implications for mixing and cooling processes. Using numerical simulations, we systematically varied viscosity, overdensity, and Mach numbers to study the transition from turbulent to laminar regimes and characterized the interplay between viscosity and radiative cooling. Our key findings can be summarized as follows:
\begin{itemize}
    \item We confirm previous results \citep[e.g.,][]{Roediger_2013, Marin-Gilabert_2022}  stating that in adiabatic setups, the KHI is suppressed when the dynamical viscosity exceeds a critical threshold. We derived an analytical solution based on the comparison between the viscous timescale and the instability growth time. We  numerically confirmed that the critical viscosity depends linearly on overdensity.

    \item At high Mach numbers in compressive-adiabatic setups, the KHI transitions from a viscosity-dominated suppression to a compressibility-induced stabilization. We verified numerically that the instability is fully suppressed above a critical Mach number, in agreement with a previous analytical work \citep{Mandelker_2016}.

    \item In a multiphase system, viscosity alters the turbulent cascade, modifying the mixing layer properties. In the weak cooling regime, viscosity suppresses turbulence, transitioning the mixing layer into a laminar state dominated by diffusion rather than convection (leading to an altered $u'-Q$ relation). Nevertheless, because cooling is the bottleneck in such a case, the overall luminosity remains unchanged.

    \item We find (perhaps surprisingly) that in the strong cooling regime,  all the TRML simulations have a similar (same order of magnitude) surface brightness and net cooling rates as inviscid simulations, regardless of the amount of viscosity. This counterintuitive behavior arises from cooling sharpening the temperature profile, thereby directly compensating for the smoothing effect of viscosity, allowing for KHI to develop, and keeping the scaling relations in TRMLs intact. 
    We derived an analytical estimate for when viscosity does not affect cooling rates based on the relative timescales of turbulence, viscosity, and cooling. Our numerical results match this theoretical prediction, confirming that cooling can compensate for the viscous suppression in sufficiently strong cooling conditions.

    \item The suppression of turbulence only mildly affects the energy dissipation in TRMLs. In the weak cooling regime, viscosity reduces turbulent velocities, but it has a weak effect on the surface brightness. In the strong cooling regime, the surface brightness increases up to a factor of $\sim 2$ with viscosity, suggesting that the smoother temperature gradient leads to a larger amount of intermediate temperature gas, thereby enhancing cooling efficiency.
\end{itemize}

These results quantify the impact of viscosity in astrophysical TRMLs,  with important implications for the dynamics of ram-pressure stripping galaxies and galaxy halos. In addition, it also provides key insights into the dynamics of the CGM and ICM.

\begin{acknowledgements}
TM would like to thank Hitesh Das, Eugene Churazov, Elke Roediger, Rajsekhar Mohapatra and Klaus Dolag for the intense discussions that motivated this work. The authors also want to thank the referee for their very useful comments. TM acknowledges support by the COMPLEX project from the European Research Council (ERC) under the European Union’s Horizon 2020 research and innovation program grant agreement ERC-2019-AdG 882679. MG thanks the Max Planck Society for support through the Max Planck Research Group, and the European Union for support through ERC-2024-STG 101165038 (ReMMU).
Most computations were performed on the HPC system Freya at the Max Planck Computing and Data Facility (MPCDF).
This research was supported in part by grant NSF PHY-2309135 to the Kavli Institute for Theoretical Physics (KITP).
\end{acknowledgements}

\bibliography{Bibtex}{}
\bibliographystyle{aa}

\appendix

\section{Growth of the KHI} \label{app:growth_KHI}

\begin{figure}
   \centering
   \includegraphics[width=\hsize]{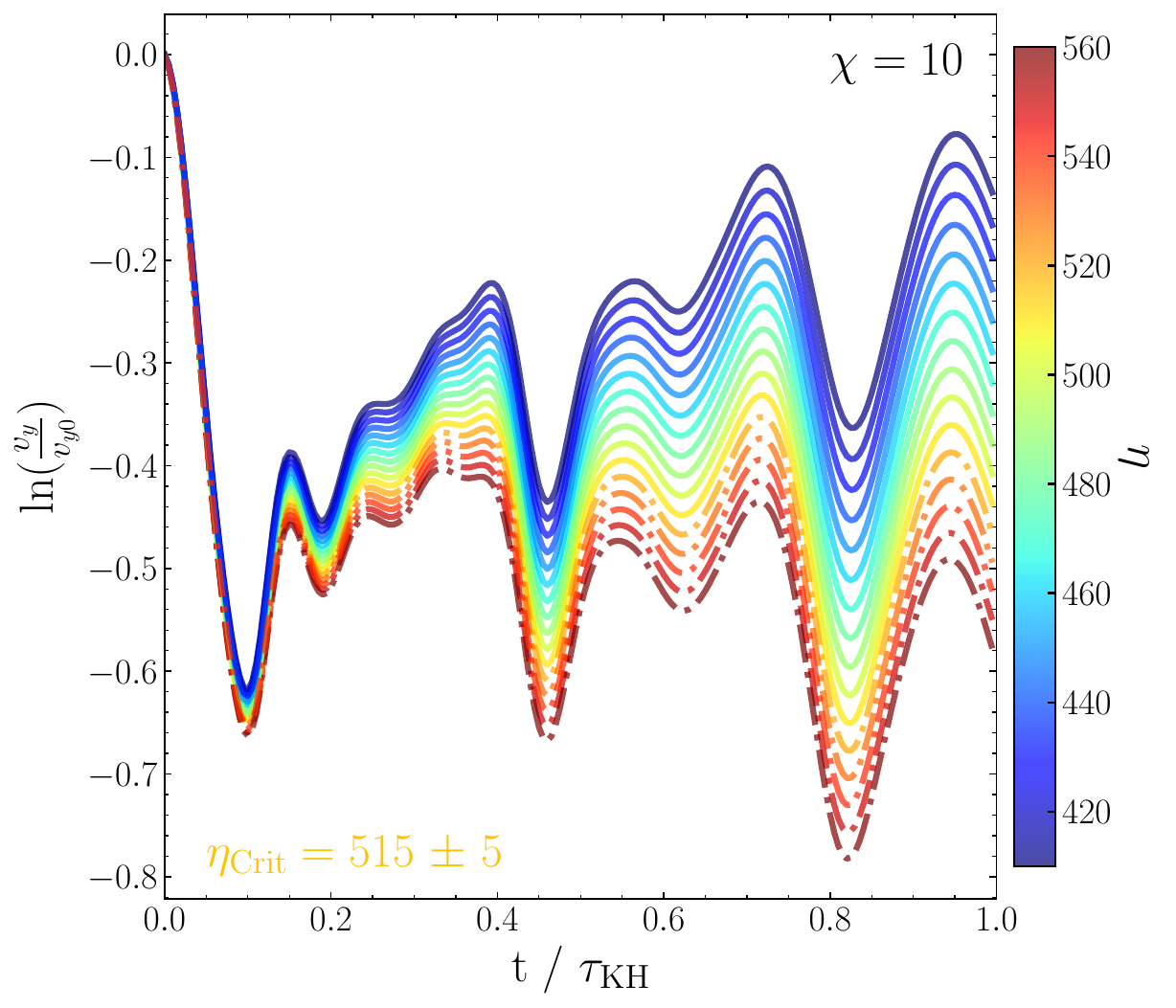}
      \caption{Growth of the instability as a function of time for an overdensity of $\chi = 10$, color-coded for different values of viscosity. The value of the critical viscosity is equal to $\eta_{\mathrm{Crit}} = 515 \pm 5$, where the error is given by the average between the viscous run with a positive and a negative slope. The solid lines indicate the simulations in which the KHI is still unstable and the dash-dotted line where the instability is suppressed ($\eta > \eta_{\mathrm{Crit}}$).}
      \label{fig:crit_visc_10chi}
\end{figure}

The growth or decay of the KHI is calculated using a discrete convolution of the sinusoidal perturbation \citep[][]{McNally_2012} to measure the rate of growth of the KHI. A positive slope within 1$\tau_{\rm KH}$ is considered a growth, while a negative slope a decay. The numerical value of critical viscosity will be the value found between the run with the shallowest positive slope and the one with the shallowest negative slope. Figure~\ref{fig:crit_visc_10chi} illustrates the measurement of the critical viscosity for an overdensity of 10, where we obtain a critical viscosity of $\eta_{\rm Crit}=515\pm5$ in code units. The initial decay is due to the loss of kinetic energy of the fluid moving perpendicular to the shear flow through the flow moving in the opposite direction \citep[][]{Junk_2010}.

\section{Mass conversion and luminosity} \label{app:mdot_L}

\begin{figure}
   \centering
   \includegraphics[width=\hsize]{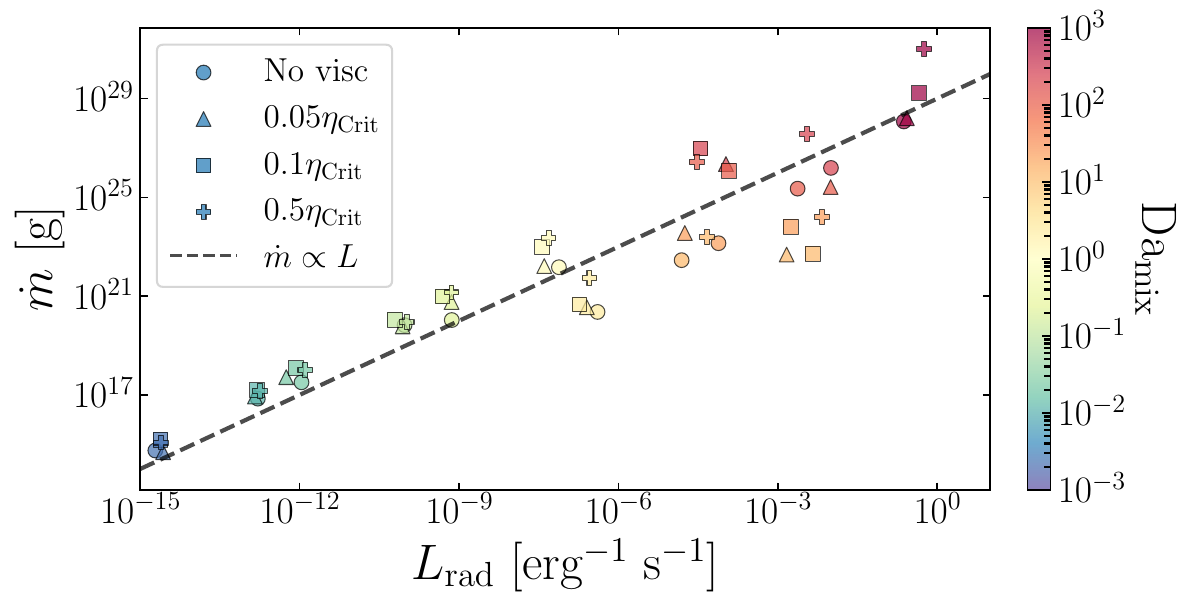}
      \caption{Mass conversion from hot to cold gas as a function of the total luminosity, color-coded by $\rm Da$. The mass transfer is proportional to the total luminosity, indicating that the luminosity is due to enthalpy flux.}
      \label{fig:mdot_L}
\end{figure}

In radiative mixing layers, the net radiative losses due to cooling are balanced by the enthalpy flux, producing a mass flow from the hot to the cold medium. This means that the total luminosity is proportional to the mass transfer from the hot to the cold medium: $\dot{m} \propto L_{\rm rad}$ \citep{Ji_2019, Gronke_2022}. We checked that the total luminosity measured in our system was due to enthalpy flux and not to dissipation of kinetic energy by measuring the mass transfer from hot to cold gas and the total luminosity (see Fig.~\ref{fig:mdot_L}). The linear dependence (dashed black line) shows that the luminosity in our system is due to enthalpy flux (mass transfer from hot to cold gas), and not due to dissipation of kinetic energy or viscous heating, regardless of the amount of viscosity and the $\rm Da$.

\section{Intermediate temperature histogram} \label{app:temp_hist}

\begin{figure*}
   \centering
   \includegraphics[width=0.95\hsize]{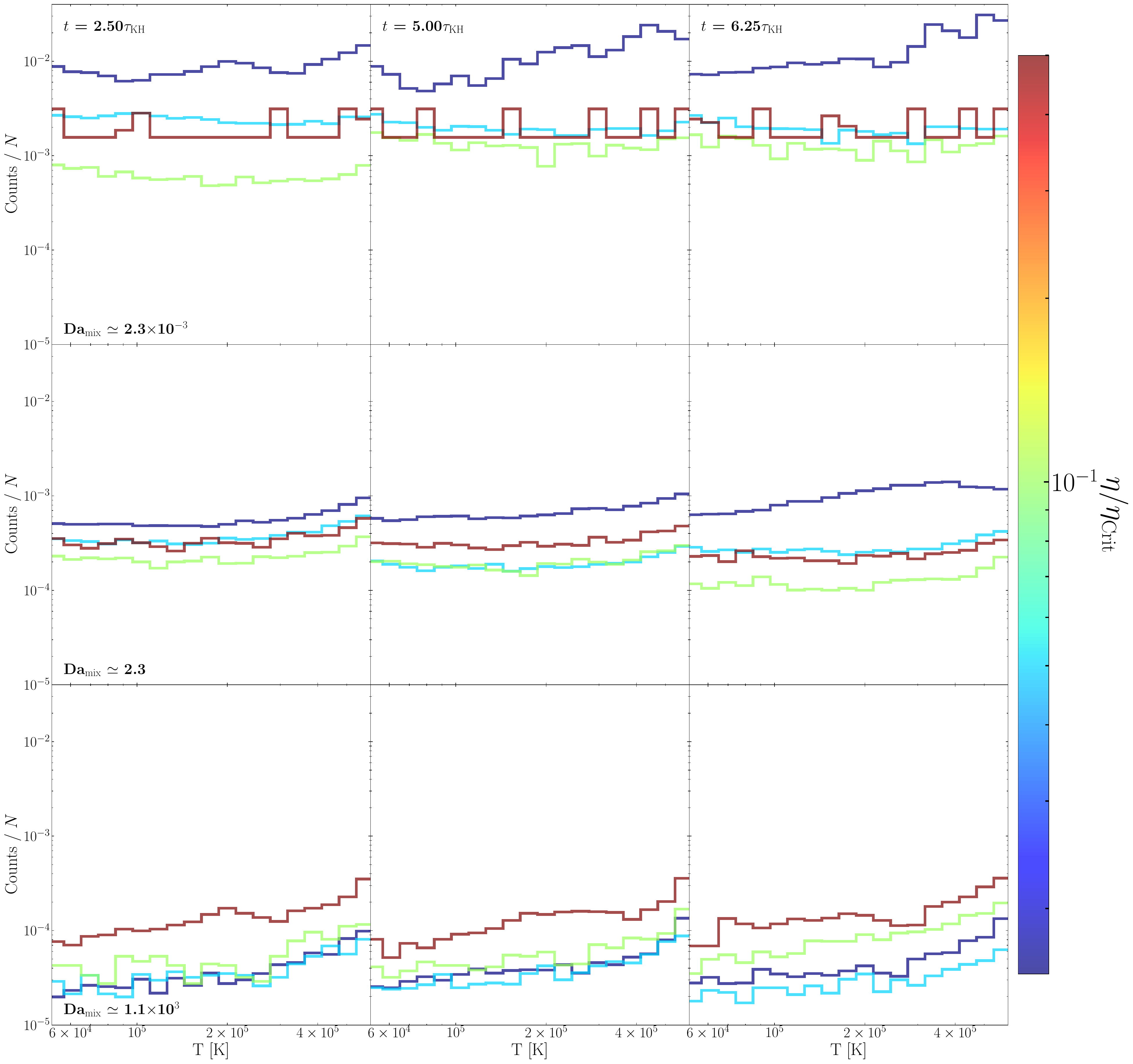}
      \caption{Intermediate temperature histograms color-coded by viscosity. \textit{From top to bottom}:  Weak cooling ($\rm Da_{mix} \simeq2.3\times10^{-3}$), intermediate cooling ($\rm Da_{mix} \simeq2.3$), and strong cooling ($\rm Da_{mix} \simeq1.1\times10^{3}$). \textit{From left to right:} Histogram measured at $t = 2.5\tau_{\rm KH}$, $t = 5\tau_{\rm KH}$ and $t = 6.25\tau_{\rm KH}$.}
      \label{fig:temp_hist}
\end{figure*}

The cooling curve reaches the highest values at intermediate temperatures, making cooling more effective in this range. This means that the amount of $Q$ for a given $\rm Da$ depends on the amount of gas with an intermediate temperature. Quantifying the amount of intermediate gas gives us a hint on how $Q$  responds in different simulations. Figure~\ref{fig:temp_hist} shows the histogram of intermediate temperature gas for different $\rm Da_{mix}$ (top to bottom: $\rm Da_{mix} \simeq2.3\times10^{-3}$, $\rm Da_{mix} \simeq2.3$ and $\rm Da_{mix} \simeq1.1\times10^{3}$) and different times (left to right: $t = 2.5\tau_{\rm KH}$, $t = 5\tau_{\rm KH}$ and $t = 6.25\tau_{\rm KH}$).

In the weak cooling regime (top row), the "No visc"\ run mixes the gas more effectively, leading to a larger amount of intermediate temperature gas. However, since cooling is the bottleneck in this case, this does not make a big impact on $Q$ (see Fig.~\ref{fig:u_turb_vs_Q}). In the viscous runs, the amount of intermediate temperature gas is very similar. This is due to the smooth momentum profile as a consequence of viscosity. Due to pressure equilibrium, this smoother momentum is translated into a smoother temperature profile, leading to more intermediate temperature gas. Note the flat histogram in the case of $0.5\eta_{\rm Crit}$ due to its laminar flow, where the temperatures are distributed approximately equally along the $y$-axis. For $\rm Da_{mix} \simeq2.3$ (middle row), the histograms look very similar, since at this value of $\rm Da_{mix}$ all the simulations are already cooling-dominated, although at later times the "No visc" run seems to produce more intermediate temperature gas. In the strong cooling regime (bottom row), viscosity leads to a slightly broader temperature profile, and therefore more intermediate temperature gas. Although the effect is only a factor of $\sim 2$, the strong cooling produces that this effect can be visible when $Q$ is calculated (see the right panel of Fig.~\ref{fig:Q_comparison}).

\section{Shear velocity profiles} \label{app:vel_prof}

\begin{figure}
   \centering
   \includegraphics[width=\hsize]{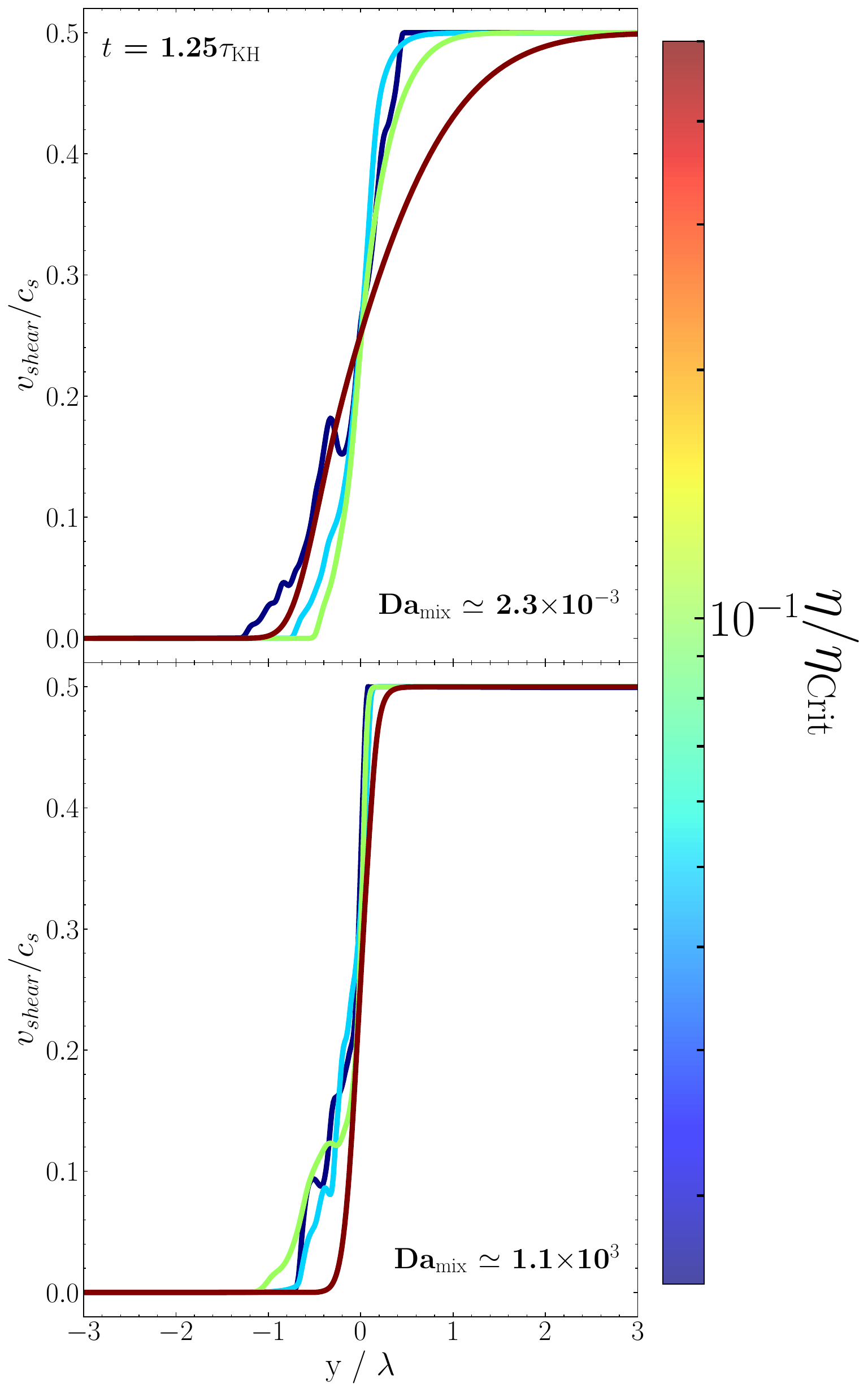}
      \caption{$v_{\rm shear}$ profiles after $t = 1.25\tau_{\rm KH}$ color-coded by viscosity. \textit{Top:} Results for $\rm Da_{mix} \simeq2.3\times10^{-3}$, where the velocity gradient is smoothed depending on the amount of viscosity. \textit{Bottom:} Results for $\rm Da_{mix} \simeq1.1\times10^{3}$, where cooling dominates over viscosity and keeps the profiles sharp.}
      \label{fig:vx_profile}
\end{figure}

In a planar slab (or sheet) setup, the effect of viscosity in the velocity gradient is obtained by solving the Rayleigh problem \citep[also known as Stokes' first problem;][]{Stokes_1851, Rayleigh_1911}:
\begin{equation}
    v_{\rm Shear}(y) = |v_{\rm Shear_0}| \, \mathrm{erf} \, \left(\frac{y}{2 \sqrt{\nu t}} \right) \, .
    \label{eqn:error_function}
\end{equation}
Here, $v_{\rm Shear_0}$ is the initial shear velocity, $y$ is the direction perpendicular to the interface, $t$ is the time and $\nu$ is the kinematic viscosity. The more viscous the medium is, the faster the velocity gradient is smoothed out. This can be seen in the top panel of Fig.~\ref{fig:vx_profile}, where in the weak cooling regime a higher viscosity leads to a smoother velocity profile within the same amount of time. The effect of turbulence can be seen in the "No visc"\ case, where the profile is not completely smooth. However, in the strong cooling regime all the profiles remain sharp after the same amount of time due to the effect of cooling. 

\section{Measurement of the smoothing distance, $d$} \label{app:d_shear}

\begin{figure}
   \centering
   \includegraphics[width=\hsize]{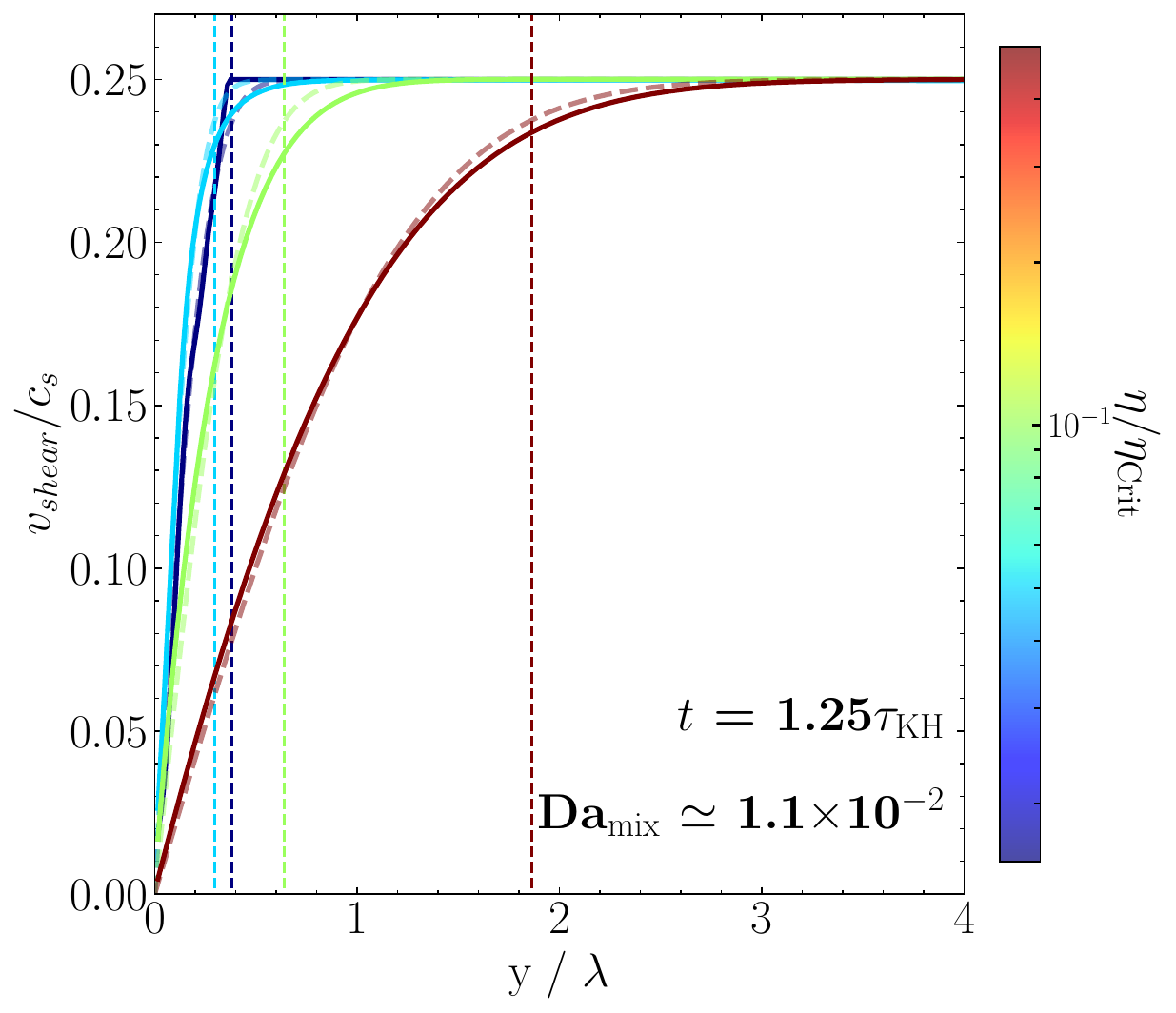}
      \caption{$v_{\rm shear}$ profiles at $\rm Da_{mix} \simeq2.3\times10^{-3}$ and $t = 1.25\tau_{\rm KH}$ showing the fit of Eq. \ref{eqn:error_function} for the different viscosities (thick dashed lines). The distance $d$ at which the shear profile has been smoothed 5\% with respect to the original value is given by the vertical dashed lines.}
      \label{fig:vx_profile_fit}
\end{figure}

To quantify the distance, $d$, at which the velocity gradient has been smoothed due to the effect of viscosity, we focus on the hot medium ($y > 0$; see Fig.~\ref{fig:vx_profile_fit}). The reason for this is that, in a system with a constant dynamic viscosity ($\eta$), the high density of the cold medium leads to a lower diffusion coefficient ($\nu$), therefore the overall diffusion of the system is dominated by the hot gas (see Sect. \ref{sec:hot_vs_cold_viscosity}). The effect of viscosity is given by Eq. \ref{eqn:error_function}, which tells us how much the velocity gradient has been smoothed after a time $t$ for a given viscosity. By fitting this equation to our data (thick dashed lines), we have a smooth function for our system, avoiding spurious artifacts that can affect the results. Once we have fitted the function, we calculate the distance $d$ from the center to the point where the velocity gradient has been smoothed 5\% with respect to the initial value (vertical dashed lines).

\section{Convergence test}\label{sec:convergence}

To make sure that our results are robust against numerical resolution, we have performed a sample of simulations for the weak and strong cooling regimes ${\rm Da_{mix}} \simeq 2.3 \times 10^{-3}$ and ${\rm Da_{mix}} \simeq 1.1 \times 10^{3}$, respectively, varying the resolution: from ($16\times160\times16$) to ($128\times1280\times128$). 

\begin{figure}
    \centering
        \includegraphics[width=\hsize]{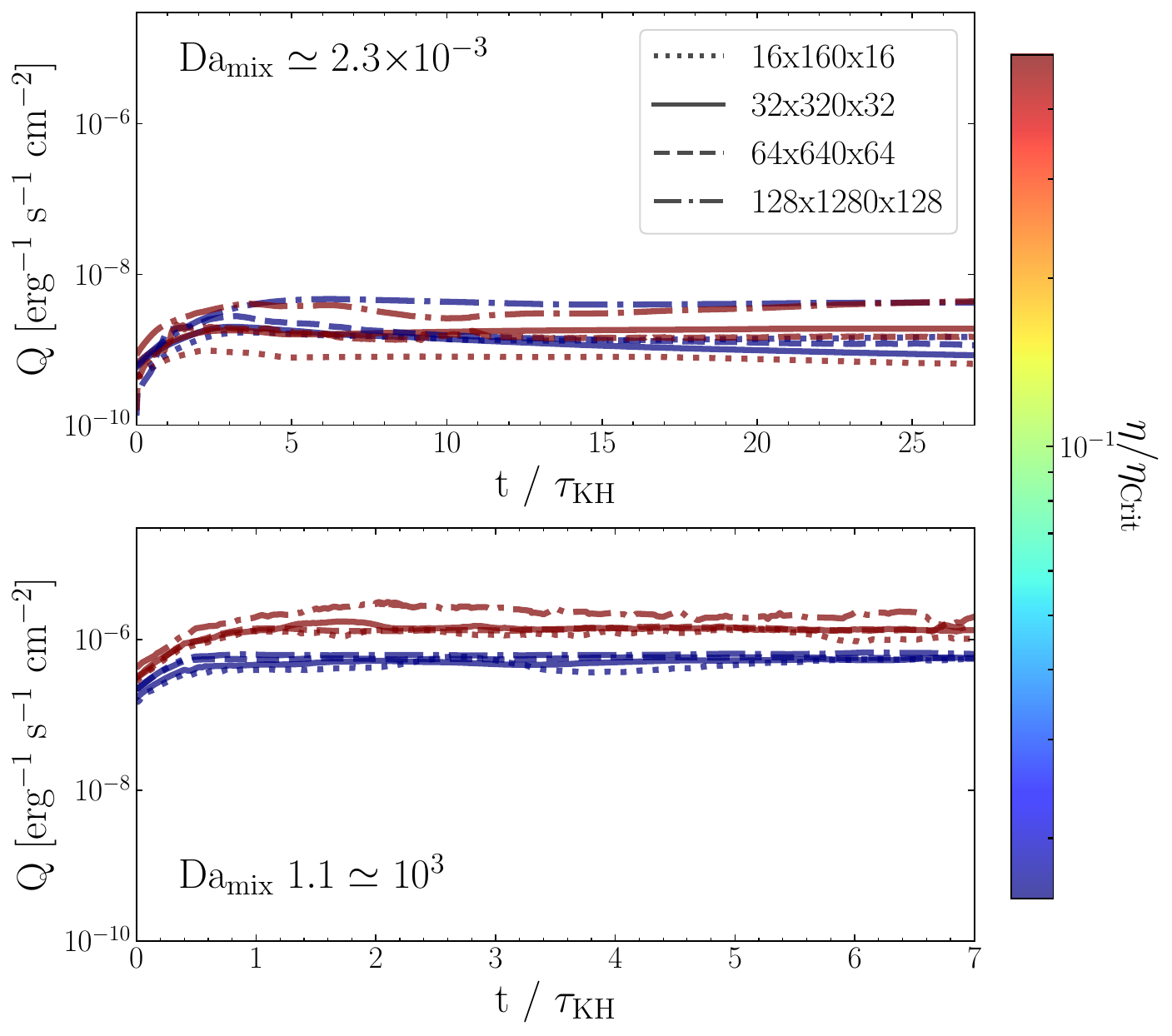}
    \caption{Evolution of the surface brightness due to cooling varying resolution from ($16\times160\times16$) to ($128\times1280\times128$) for the nonviscous (blue) and $0.5\eta_{\rm Crit}$ (red) cases. {\it Top}: Weak cooling regime, ${\rm Da_{mix}} \simeq 2.3 \times 10^{-3}$. {\it Bottom}: Strong cooling regime, ${\rm Da_{mix}} \simeq 1.1 \times 10^{3}$.}
    \label{fig:Q_comparison_res}
\end{figure}

Figure \ref{fig:Q_comparison_res} shows the evolution of $Q$ over time for the weak cooling regime (top panel) and for the strong cooling regime (bottom panel), for the nonviscous (blue) and $0.5\eta_{\rm Crit}$ (red) cases at different resolutions. In all cases, the results shown in the paper hold: no significant difference between viscous and inviscid cases in the weak cooling regime, and a small increase (factor of $\sim2$) in the strong cooling regime. It is important to note two details:
\begin{itemize}
    \item The runs with the highest resolution in the weak cooling regime lead to a slightly larger surface brightness compared to the runs with lower resolution. However, far from making it worse, an increase in $Q$ means that the data points in Fig.~\ref{fig:Q_vs_Da} of the paper become closer to the expected value at that given Da.
    \item The viscous run with the highest resolution in the strong cooling regime leads to a slight increase in $Q$. We attribute this to numerical noise associated with the frame boost (see Sect. \ref{sec:3D_setup}) that we found when increasing resolution, and not a lack of convergence (note that the rest of the runs are perfectly converged).
\end{itemize}

Another important note here is that, since we have changed the resolution, as a consequence, we have modified the numerical conduction. This means that, in this analysis, the Karlovitz number is expected to be different than the one given throughout the whole paper. 
Changing both the viscosity and conduction in a controlled manner, thus, assessing the impact of mixing and cooling in the full Ka range, is a logical next step but beyond the scope of the current study.

\end{document}